\documentclass[aps,twocolumn,preprintnumbers,amsmath,amssymb,floatfix,longbibliography]{revtex4-1}
\usepackage{graphicx}
\usepackage{graphics}
\usepackage{psfrag}
\usepackage{hyperref}
\usepackage{multirow}
\usepackage[sort&compress]{natbib}
\usepackage{amssymb}
\usepackage{amsmath}
\usepackage{amsthm}
\usepackage{bbold}
\usepackage{bbm}
\usepackage{enumerate}

\newcommand{\vare}{\varepsilon}

\newcommand{\rmi}{{\rm i}}

\newcommand{\vphi}{\varphi}

\usepackage{xcolor}

\begin{document}

\hypersetup{pdftitle={Critical phenomena at the complex tensor ordering phase transition}}
\title{Critical phenomena at the complex tensor ordering phase transition}

\author{Igor Boettcher}
\affiliation{Department of Physics, Simon Fraser University, Burnaby, British Columbia, V5A 1S6, Canada}
\author{Igor F. Herbut}
\affiliation{Department of Physics, Simon Fraser University, Burnaby, British Columbia, V5A 1S6, Canada}

\begin{abstract}
We investigate the critical properties of the phase transition towards complex tensor order that has been proposed to occur in spin-orbit coupled superconductors. For this purpose we formulate the bosonic field theory for fluctuations of the complex irreducible second-rank tensor order parameter close to the transition. We then determine the scale dependence of the couplings of the theory by means of the perturbative Renormalization Group (RG). For the isotropic system we generically detect a fluctuation-induced first-order phase transition. The initial values for the running couplings are determined by the underlying microscopic model for the tensorial order. As an example we study three-dimensional Luttinger semimetals with electrons at a quadratic band touching point. Whereas the strong-coupling transition of the model receives substantial fluctuation corrections, the weak-coupling transition at low temperatures is rendered only weakly first-order due to the presence of a fixed point in the vicinity of the RG trajectory. If the number of fluctuating complex components of the order parameter is reduced by cubic anisotropy, the theory maps onto the field theory for frustrated magnetism.
\end{abstract}

\maketitle

\section{Introduction}

Recent years have witnessed a revolution in the synthesis and study of three-dimensional superconducting materials with multiband touching points in the electronic band dispersion. The modified nature of the electronic degrees of freedom can lead to exotic or topological superconducting states \cite{RevModPhys.63.239,SmidmanReview}. An example is provided by the half-Heusler superconductors RPtBi and RPdBi with R a rare-earth like atom \cite{Chadov,yan}, where strong spin-orbit coupling due to the heavy element Bi induces band inversion and quadratic band touching (QBT) near the Fermi energy so that the electronic degrees of freedom carry an effective spin $3/2$ \cite{abrikosov,luttinger,PhysRevB.69.235206,moon,PhysRevLett.113.106401,BalentsReview} and many Cooper pairing channels with spins ranging from $0$ to $3$ are possible and have been studied theoretically \cite{PhysRevLett.116.177001,PhysRevB.93.205138,PhysRevLett.118.127001,2016arXiv160303375K,PhysRevB.95.075149,2017arXiv170702739T,2017arXiv170703831S,BoettcherComplex,PhysRevB.96.144514,2017arXiv170807825R,2017arXiv170904487V,2017arXiv170904487V,2017arXiv171008428M}. Numerous types of other higher-pseudospin fermions have been proposed to describe multiband touching points and make this direction of research on superconductivity particularly rich \cite{PhysRevB.92.081304,PhysRevB.93.045113,PhysRevB.93.241113,PhysRevB.94.195205,2017arXiv170101553K,PhysRevLett.119.206402}.

For three-dimensional systems with an inverted QBT point, which we refer to as Luttinger semimetals, suitable electron-electron interactions can induce a superconducting complex tensor order state describing a condensate of Cooper pairs with spin $2$ \cite{BoettcherComplex}. The corresponding order parameter is given by a complex irreducible second-rank tensor, which can be represented by a symmetric traceless $3\times 3$ complex matrix. The mean-field phase structure of the model shows both first- and second-order transitions. In particular, at weak coupling, the transition is of second order at the mean-field level, and the corresponding critical phenomena may thus be observable in the half-Heusler superconductors. 

In this work, we address the question of the critical properties of the complex tensor ordering phase transition. For this we construct the bosonic field theory for the order parameter close to the critical point and determine the running of couplings by means of the perturbative renormalization group (RG). We find that fluctuations of the complex tensor field significantly influence the nature of the phase transition. We emphasize that the matrix field theory considered here is distinct from matrix models considered in the context of superfluid He-3 \cite{BookVolovik} or high energy physics \cite{PhysRevD.29.338,PhysRevB.23.3549,BERGES1997675}, because in these cases the fluctuating matrix field is not restricted to be symmetric and traceless. The physics discussed here may also be observed in spin-2 Bose--Einstein condensates of ultracold atoms \cite{PhysRevA.65.063602,0305-4470-36-32-302,0034-4885-77-12-122401,2017arXiv171101155T}.

Here we summarize our main findings. The free energy for the complex tensor $\phi$ close to the second-order transition is given by
\begin{align}
 \nonumber F(\phi) =&{} \frac{r}{2}\mbox{tr}(\phi^\dagger\phi) + \frac{\bar{\lambda}_1}{4}[\mbox{tr}(\phi^\dagger\phi)]^2 \\
 \label{intro1} &+\frac{\bar{\lambda}_2}{4} |\mbox{tr}(\phi^2)|^2 + \bar{\lambda}_3 \mbox{tr}(\phi^\dagger\phi\phi^\dagger\phi)
\end{align}
with $r\propto (T-T_{\rm c})$. The symmetry group of the theory is $\text{SO}(3)\times \text{U}(1)$. Note that a term $\mbox{tr}(\phi^3)$, characteristic for real tensor order describing liquid crystals, is forbidden here due to the global $\text{U}(1)$ symmetry. For $r<0$ and generic values of the couplings $\{\bar{\lambda}_m\}$, the ground state of the theory is determined from Eq. (\ref{intro1}). However, for $\bar{\lambda}_3=0$ the expression has an accidental $\text{SO}(5)\times \text{U}(1)$ symmetry \cite{PhysRevA.9.868} so that the ground state is determined by additional terms of sextic order in the field \cite{BoettcherComplex}. At the critical point ($r=0$), fluctuations of the order parameter lead to a scale dependence of the couplings $\{\bar{\lambda}_m\}$ and can thereby alter the equilibrium state. To one-loop order we find a fluctuation-induced first-order transition.

For the complex tensor ordering transition in isotropic Luttinger semimetals we have $\bar{\lambda}_3=0$ at the mean-field level. Due to the enlarged $\text{SO}(5)\times\text{U}(1)$ symmetry of the quartic free energy in this case, one might expect at first sight that a coupling $\bar{\lambda}_3\neq0$ cannot be generated by the running of $\bar{\lambda}_1$ and $\bar{\lambda}_2$. However, the field theory for complex tensor order features a nonstandard kinetic term parametrized by a parameter $K$ that reduces the symmetry to the physical $\text{SO}(3)\times\text{U}(1)$. Consequently, the coupling $\bar{\lambda}_3$ is generated during the RG flow for $K\neq 0$. We find $K\sim 1$ in Luttinger semimetals at the mean-field level, so that the accidental degeneracy of the quartic theory is lifted beyond mean-field theory.

In superconducting materials such as the half-Heusler compounds, rotation symmetry is reduced by the cubic anisotropy of the crystal structure. This effect can suppress fluctuations of the complex tensor order parameter and thereby modify the critical properties. In particular, whereas the complex tensor in the isotropic case is described by $N=5$ fluctuating complex components, cubic symmetry can reduce the number of fluctuating complex components to $N=2,3$. For instance, for YPtBi \cite{PhysRevB.84.220504,PhysRevB.86.064515,PhysRevLett.116.137001,2016arXiv160303375K} we deduce $N=2$ from the electronic band structure \cite{2016arXiv160303375K}. Under certain conditions on the parameters of the theory that we specify below, the phase transition is then captured by the $\text{O}(N)\times \text{O}(2)$ symmetric theory for frustrated magnetism, which also describes the transition of stacked triangular antiferromagnets, helimagnets, or the dipole-locked A-phase of He-3, see Refs. \cite{PelissettoVicari,PhysRevB.69.134413} for an overview. The phase structure of this model is highly controversial \cite{Kawamura1,PhysRevB.38.4916,1995PhLA..208..161A,PhysRevB.63.140414,2001NuPhB.607..605P,PelissettoVicari,PhysRevB.69.134413,PhysRevA.93.051603,PhysRevA.94.053623}, with contradicting conclusions from perturbative RG, nonperturbative RG, Monte Carlo computations, and experiment. In particular, it has been conjectured that the transition lies in an exotic chiral universality class \cite{Kawamura1,PhysRevB.38.4916} or that it is of first order. A detailed discussion of different scenarios is presented below.

This work is organized as follows. In Sec. \ref{SecField} we introduce the complex tensor field theory that captures critical phenomena close to a second-order or weak first-order phase transition. In Sec. \ref{SecRG} we then analyze the running of couplings of the theory due to fluctuations of the order parameter. In Sec. \ref{SecQBT} we elaborate the relevance of our findings for superconductivity in Luttinger semimetals. In Sec. \ref{SecDis} we summarize our results and point out directions for future work. In the Appendices we present our convention for the representation of Gell-Mann matrices, clarify the difference to matrix models with $\text{U}(3)\times\text{U}(3)$ symmetry, derive inequalities for tensor invariants, compute the RG beta functions that are used in the main text, and list the initial conditions for the flow that describes the transition in Luttinger semimetals.

\section{Field theory of complex tensor boson}\label{SecField}

In the following we construct the field theory that captures fluctuations of a complex irreducible second-rank tensor boson close to a second- or weak first-order phase transition. We first introduce our notation for the two key parametrizations of the fluctuating field, then derive the Lagrangian for the theory close to the phase transition, discuss the stable parameter regime of the model together with how to physically interpret its instability, and eventually present the modification of the model due to cubic anisotropy.

\subsection{Field representations}

The irreducible second-rank complex tensor field can be represented in several equivalent ways. In particular, we can identify it with a complex symmetric traceless $3\times 3$ matrix $\phi_{ij}$ ($i,j=1,2,3$) or its five independent complex components $\vphi_a$ ($a=1,\dots,5$). We employ both parametrizations interchangeably, using whichever makes our arguments most transparent. Here we briefly summarize the most important relations to translate between both pictures.

The irreducibility of the tensor $\phi$ is equivalent to being symmetric and traceless, $\phi_{ij}=\phi_{ji}$ and $\delta_{ij}\phi_{ij}=0$. Under $R\in\text{SO}(3)$ it transforms as $\phi_{ij}\to R_{ik}R_{jl}\phi_{kl}$. We can represent $\phi$ by a symmetric traceless matrix that transforms as
\begin{align}
 \label{field1b} \phi\to R\phi R^{\rm T}.
\end{align}
Two configurations $\phi$ and $\phi'$ are physically equivalent if there exists an $R$ such that $\phi' = e^{\rmi \alpha} R\phi R^{\rm T}$ with $e^{\rmi \alpha}$ a phase. The matrix $\phi$ can be parametrized as 
\begin{align}
 \label{field1} \phi = \begin{pmatrix} \vphi_1-\frac{1}{\sqrt{3}}\vphi_2 & \vphi_5 & \vphi_3 \\ \vphi_5 & -\vphi_1-\frac{1}{\sqrt{3}} \vphi_2 & \vphi_4 \\ \vphi_3 & \vphi_4 & \frac{2}{\sqrt{3}}\vphi_2 \end{pmatrix}.
\end{align}
Here $\{\vphi_a\}$ are the five independent complex components of the tensor. We have
\begin{align}
\label{field2} \phi_{ij} = \vphi_a \Lambda^a_{ij}
\end{align}
with the five real Gell-Mann matrices $\{\Lambda^a\}$ given in Eq. (\ref{field3}). They form an orthogonal basis for symmetric traceless $3\times 3$ matrices with orthogonality relation $\mbox{tr}(\Lambda^a\Lambda^b) = 2\delta_{ab}$. More generally we define the $J$-symbols
\begin{align}
 \label{field4} J_{ab\dots c} = \mbox{tr}(\Lambda^a\Lambda^b \dots \Lambda^c).
\end{align}
Obviously, $J_a=0$ and $J_{ab}=2\delta_{ab}$. One easily verifies that $J_{abc}$ is invariant under any permutation of its indices.

We collect the components of $\phi$ in the five-tuple
\begin{align}
 \label{field5} \vec{\vphi}=(\vphi_1,\vphi_2,\vphi_3,\vphi_4,\vphi_5)^{\rm T}\in \mathbb{C}^5,
\end{align}
which transforms under the five-dimensional ($\ell=2$) representation of $\text{SO}(3)$: For $R\in\text{SO}(3)$ we have $ \vphi_a \to M_{ab}(R) \vphi_b$ with $M_{ab}(R) = \frac{1}{2}\mbox{tr}(R\Lambda^b R^{\rm T}\Lambda^a)$. We denote $\vec{\vphi}$ with a vector arrow and employ the notation
\begin{align}
 \label{field6b} |\vec{\vphi}|^2 := \vphi_a^*\vphi_a,\ \vec{\vphi}^2 := \vphi_a\vphi_a.
\end{align}

\subsection{Effective action}

In this section we construct the most general low-energy effective action for a complex tensor which is invariant under $\text{SO}(3)\times \text{U}(1)$ transformations. In order to address the physics close to the transition it is sufficient to consider the long-wavelength limit and the limit of small field amplitudes. We thus restrict the effective action to terms of second order in derivatives of $\phi$ and to terms of at most fourth order in $\phi$. We show that the most general such action compatible with the symmetry group is then given by
\begin{align}
 \nonumber S = \int\mbox{d}^3x \Biggl[{}& \vphi_a^*\Bigl(-\delta_{ab}\nabla^2+\frac{K}{\sqrt{3}}J_{abc}d_c(-\rmi \nabla)\Bigr)\vphi_b \\
 \label{flow0} &+ \bar{\lambda}_1|\vec{\vphi}|^4+\bar{\lambda}_2|\vec{\vphi}^2|^2+\bar{\lambda}_3\mbox{tr}(\phi^\dagger\phi\phi^\dagger\phi)\Biggr],
\end{align}
with the $d_a$-functions defined in Eq. (\ref{field11}) below. The upper critical dimension for this theory is four. For our derivation we write
\begin{align}
 \label{field0} S = \int \mbox{d}^3 x (L_{\rm kin} + L_{\rm int}),
\end{align}
where $L_{\rm kin}$ and $L_{\rm int}$ constitute the kinetic and interaction parts of the effective Lagrangian, respectively. Remarkably, complex tensor order in three dimensions features peculiar properties in both terms.

The kinetic part of the Lagrangian contains the contribution
\begin{align}
\label{field8} \mbox{tr}(\nabla \phi^\dagger\cdot\nabla \phi) = (\partial_i \phi_{jk}^*)(\partial_i \phi_{jk}) =2 (\nabla\vphi_a^* \cdot\nabla \vphi_a)
\end{align}
that is commonly encountered in bosonic field theories. However, in three spatial dimensions also the term
\begin{align}
 \label{field9} (\partial_i \phi_{ik}^*)(\partial_j \phi_{jk}) = (\partial_i \vphi_a^*)(\partial_j\vphi_b) (\Lambda^a\Lambda^b)_{ij}
\end{align}
is compatible with $\text{SO}(3)\times \text{U}(1)$ symmetry. Note that this term is reminiscent of the term $(\nabla\cdot \vec{m})^2$ that is admissible for the three-dimensional Heisenberg model, or $(\partial_i T_{ik})(\partial_j T_{jk})$ for nematic order described by a real second-rank tensor $T$ \cite{PhysRevB.92.045117}.

We introduce the functions $d_a(\textbf{p}) = \frac{\sqrt{3}}{2}p_i p_j \Lambda^a_{ij}$ so that
\begin{align}
 \nonumber d_1(\textbf{p}) &= \frac{\sqrt{3}}{2}(p_x^2-p_y^2),\ d_2(\textbf{p}) = \frac{1}{2}(2p_z^2-p_x^2-p_y^2),\\
 \label{field11} d_3(\textbf{p}) &= \sqrt{3}p_zp_x,\ d_4(\textbf{p}) = \sqrt{3}p_yp_z,\ d_5(\textbf{p}) = \sqrt{3}p_xp_y.
\end{align}
The normalization is such that $\sum_{a=1}^5d_a(\textbf{p})^2=p^4$. We then rewrite the right hand side of Eq. (\ref{field9}) by means of \cite{PhysRevB.95.075149}
\begin{align}
 \label{field10} p_i p_j (\Lambda^a\Lambda^b)_{ij} &= \frac{2}{3}p^2 \delta_{ab} +\frac{1}{\sqrt{3}}J_{abc} d_c(\textbf{p}).
\end{align}
Whereas the first term merely results in a shift of the prefactor of $\mbox{tr}(\nabla \phi^\dagger\cdot\nabla \phi) $, the second contribution is genuinely different. We can thus parametrize the most general kinetic part to the Lagrangian in three dimensions with up to two derivatives as
\begin{align}
 \label{field12} L_{\rm kin} = \vphi_a^*\Bigl( -\delta_{ab}\nabla^2 + \frac{K}{\sqrt{3}} J_{abc}d_c(-\rmi \nabla)+r\delta_{ab}\Bigr)\vphi_b,
\end{align}
where $K$ is a dimensionless parameter. We further included a quadratic mass-like term.

In order to construct the interaction part of the Lagrangian, we note that terms cubic in the field $\phi$ are forbidden due to $\text{U}(1)$ invariance. At the quartic level, the four terms that can be constructed for a symmetric traceless complex matrix are given by
\begin{align}
 \label{field13} Q_1 &= [\mbox{tr}(\phi^\dagger\phi)]^2 = 4|\vec{\vphi}|^4,\\ 
 \label{field14} Q_2 &= |\mbox{tr}(\phi^2)|^2= 4|\vec{\vphi}^2|^2,\\
 \label{field15} Q_3 &= \mbox{tr}(\phi^\dagger\phi\phi^\dagger\phi),\\
 \label{field16} Q_4 & =\mbox{tr}(\phi^\dagger{}^2\phi^2).
\end{align}
However, in the case of $\phi$ being also a $3\times 3$ matrix we additionally have the relation
\begin{align}
 \label{field17} 2Q_1+ Q_2 = 2Q_3 + 4Q_4.
\end{align}
This can be checked by direct computation, or by inserting $A=\phi+\alpha \phi^\dagger$ into $\mbox{tr}(A^4)=\frac{1}{2}[\mbox{tr}(A^2)]^2$, which is valid for any symmetric traceless $3\times 3$ \emph{complex} matrix $A$, and by reading off the coefficient of $\alpha^2$. We may thus eliminate $Q_4$ by expressing it through the remaining three quartic terms. This is a special case of the fact that $\text{SO}(3)$ invariance  of the interaction part implies that it is a polynomial in the eight invariants \cite{Matteis2008,BoettcherComplex}
\begin{align}
 \nonumber I_1 &=\mbox{tr}(\phi^\dagger\phi),\ I_2 = \mbox{tr}(\phi^2),\ I_3=\mbox{tr}(\phi^\dagger{}^2),\\
 \nonumber I_4 &=\mbox{tr}(\phi^3),\ I_5 = \mbox{tr}(\phi^\dagger{}^3),\ I_6=\mbox{tr}(\phi^2\phi^\dagger),\\
 \label{field18} I_7&=\mbox{tr}(\phi^\dagger{}^2\phi),\ I_8 = \mbox{tr}(\phi^\dagger\phi\phi^\dagger\phi).
\end{align}
We thus conclude that the interaction part of the Lagrangian including up to four powers of the field is given by
\begin{align}
 \label{field19} L_{\rm int} = \bar{\lambda}_1 |\vec{\vphi}|^4 + \bar{\lambda}_2 |\vec{\vphi}^2|^2 + \bar{\lambda}_3 \mbox{tr}(\phi^\dagger\phi\phi^\dagger\phi).
\end{align}
In our presentation we often refer to the rescaled couplings
\begin{align}
 \label{field19b} \lambda_m = \frac{\text{S}_d \Omega^{d-4}}{(2\pi)^d}\bar{\lambda}_m
\end{align} 
with dimension $d$, $\text{S}_d$ the area of the $(d-1)$-dimensional sphere, and ultraviolet momentum cutoff $\Omega$, see Sec. \ref{SecRG}.

The Lagrangian $L_{\rm kin}+L_{\rm int}$ is constructed to respect $\text{SO}(3)\times \text{U}(1)$ symmetry of an irreducible second-rank tensor field $\phi$. Depending on the parameters of the theory, however, the symmetry group can be enlarged:
\begin{itemize}
 \item[(i)] If $K=0$ and $\lambda_2=\lambda_3=0$, the theory is invariant under $\vphi_a \to M_{ab}\vphi_b$ with $M\in \text{U}(5)$. 
 \item[(ii)] If $K=0$ and $\lambda_3=0$, the theory possesses an enlarged $\text{SO}(5)\times \text{U}(1)$ symmetry with respect to $\vphi_a\to M_{ab}\vphi_b$ with $M\in\text{SO}(5)$.
\end{itemize}
Crucially, if $\vphi_a$ are the five components of a symmetric traceless matrix, then $\vphi_a'=M_{ab}\vphi_b$ with either $M\in\text{U}(5)$ or $M\in\text{SO}(5)$ are still viable components of (another) symmetric traceless matrix. Consequently, the RG flow preserves the enlarged symmetries in the cases (i) and (ii). This is not so for the following case: (iii) Set $K=0$ and $\lambda_2=0$. The theory with quartic terms $[\mbox{tr}(\phi^\dagger\phi)]^2$ and $\mbox{tr}(\phi^\dagger\phi\phi^\dagger\phi)$ is then reminiscent of the $\text{U}(3)\times \text{U}(3)$ symmetric matrix models that are well-known in the literature \cite{PhysRevD.29.338,PhysRevB.23.3549,BERGES1997675}. Indeed, the Lagrangian is invariant under the transformation $\phi' \to \phi=U \phi V^\dagger$ with $U,V\in\text{U}(3)$. However, the matrix $\phi'$ will generally not be symmetric and traceless. Thus the coupling $\lambda_2$ is not prohibited by symmetry even if initially absent, see App. \ref{AppMat} for a detailed discussion of the difference of both matrix models. Eventually, if $K\neq 0$, the symmetry of the theory is always reduced to $\text{SO}(3)\times \text{U}(1)$ due to the kinetic term. We will later see how the features discussed here manifest in the RG evolution of the running couplings of the theory.

\subsection{Stability and first-order transition}

The consistency of the theory described by the effective action $S$ requires the parameters $K,\lambda_{1,2,3}$ to satisfy certain inequalities that we derive in the following. If these stability bounds are not satisfied, $S$ needs to be replaced by a more elaborate description.

In momentum space, the kinetic part of the critical theory ($r=0$) is given by the quadratic form $S_{\rm kin} = \int \frac{d^3p}{(2\pi)^3} \vphi_a^*(\textbf{p})D_{ab}(\textbf{p})\vphi_b(\textbf{p})$ with
\begin{align}
 \label{field20} D_{ab}(\textbf{p}) = p^2 \delta_{ab}+\frac{K}{\sqrt{3}} J_{abc} d_c(\textbf{p}).
\end{align}
The long-wavelength approximation of keeping only terms quadratic in $\textbf{p}$ is a consistent assumption if and only if all eigenvalues of the $5\times 5$ matrix $D(\textbf{p})$ are strictly positive for all $\textbf{p}$. We determine the eigenvalues to be $(1+\frac{K}{3})p^2$ (doubly degenerate), $(1+\frac{2}{3}K)p^2$, and $(1-\frac{2}{3}K)p^2$ (doubly degenerate). Consequently, a stable theory needs to satisfy the condition
\begin{align}
 \label{field21} |K|<\frac{3}{2}.
\end{align}
Note that in writing Eqs. (\ref{field12}) and (\ref{field20}) we implicitly assumed the prefactor of $p^2\delta_{ab}$ to be positive so that we can normalize it to unity by a field redefinition. If this prefactor is not positive, the theory is unstable for any value of $K$. Negative directions of $D(\textbf{p})$ in momentum space would imply that the ground state is given by an inhomogeneous field configuration with some ordering wave vector $p_{*}>0$.

In a similar fashion stability of the field theory requires the quartic interaction term to be bounded from below: $L_{\rm int}(\phi) \to +\infty $ for $|\vec{\vphi}|\to \infty$. In this case, a second-order phase transition is induced by a sign-change of $r$. Since the field-dependent terms $Q_{1,2,3}$ in $L_{\rm int}$ are all positive, the stability is determined by the signs and relative sizes of the couplings $\lambda_{1,2,3}$. The corresponding stability bounds result from the inequalities
\begin{align}
 \label{field22} & 0\leq |\vec{\vphi}^2|^2 \leq |\vec{\vphi}|^4,\ \frac{4}{3}|\vec{\vphi}|^4 \leq \mbox{tr}(\phi^\dagger\phi\phi^\dagger\phi) \leq 4 |\vec{\vphi}|^4,
\end{align}
which we derive in App. \ref{AppIneq}. We summarize the stability criteria in Tab. \ref{TabStab}. If the quartic part of $S$ is not bounded from below, higher orders in the field expansion need to be included, which typically implies that the field theory describes a first-order transition for certain $r>0$. For the small-amplitude expansion to still be a valid assumption, only weak first-order transitions (with a small jump of the order parameter at the transition) can be considered.

\renewcommand{\arraystretch}{1.2}
\begin{table}[t]
\begin{tabular}{|c|c|c|c|}
\hline \hspace{1mm} $\lambda_1$\hspace{1mm} & \hspace{1mm} $\lambda_2$\hspace{1mm}  & \hspace{1mm} $\lambda_3$\hspace{1mm}  & \hspace{1mm} stability condition\hspace{1mm}  \\ 
\hline\hline $+$ & $+$ & $+$ & stable \\ 
\hline $+$ & $+$ & $-$ & $\lambda_1+4\lambda_3>0$ \\ 
\hline $+$ & $-$ & $+$ & $\lambda_1+\lambda_2+\frac{4}{3}\lambda_3>0$\\ 
\hline $+$ & $-$ & $-$ & $\lambda_1+\lambda_2+4\lambda_3>0$ \\ 
\hline $-$ & $+$ & $+$ & $\lambda_1+\frac{4}{3}\lambda_3>0$ \\ 
\hline $-$ & $-$ & $+$ & $\lambda_1+\lambda_2+\frac{4}{3}\lambda_3>0$ \\ 
\hline $-$ & $+$ & $-$& unstable \\ 
\hline $-$ & $-$ & $-$&  unstable \\ 
\hline 
\end{tabular} 
\caption{Stability of the quartic theory described by $S$ requires the interaction term to be bounded from below. In the table, the first three columns indicate a positive ($+$) or negative sign ($-$) of the corresponding coupling $\lambda_m$, and the fourth column gives the corresponding stability condition that needs to be satisfied. If a coupling vanishes, we can use the row with either $+$ or $-$ for that coupling, and check whether the condition can be fulfilled. (Both tests yield the same conclusion.) If the stability condition is violated, higher-order terms in the field need to be included in $S$.}
\label{TabStab}
\end{table}
\renewcommand{\arraystretch}{1}

\subsection{Cubic anisotropy}\label{SecCub}

The two quintessential symmetries of the theory, rotation invariance and particle number conservation, expressed through the global symmetry group $\text{SO}(3)\times \text{U}(1)$, tightly restrict the action $S$ for complex tensor theory in three dimensions. We have seen that the fluctuating second-rank tensor field is bound to have five complex components $\vphi_a$ and three quartic self-interaction constants $\lambda_{1,2,3}$. In particular, even if the symmetry-breaking ground state of the system condenses only, say, one or two of the components, the fluctuations about this state still comprise all five components. However, if the rotational symmetry group $\text{SO}(3)$ is reduced by the crystal structure to a discrete point group, the number of fluctuating components can be decreased. As we lay out in the next section, this may have drastic effects on the nature of the complex tensor ordering phase transition.

We focus here on the important case that rotations are restricted to the cubic group. We write $\phi=\phi_{\rm E}+\phi_{\rm T}$ with
\begin{align}
 \label{cub1} \phi_{\text{E}} &= \sum_{a=1,2}\eta_{a}\Lambda^a,\\
  \label{cub2} \phi_{\text{T}} &= \sum_{a=3,4,5} \chi_{a} \Lambda^a.
\end{align}
Then $\phi_{\rm E}$ is entirely diagonal and $\phi_{\rm T}$ is entirely off-diagonal. Importantly, if $\phi$ is diagonal (off-diagonal), it remains diagonal (off-diagonal) under cubic transformations.  The notation is to indicate that $\phi_{\rm E}$ and $\phi_{\rm T}$ constitute the $\text{E}_{\text{g}}$ (``E'') and  $\text{T}_{2\text{g}}$ (``T'') representations of the cubic group.

If the system is only cubic symmetric, the mass-like quadratic term $r |\vec{\vphi}|^2$ in the Lagrangian generically splits into the two contributions
\begin{align}
 \label{cub3} r_{\rm E} |\vec{\eta}|^2 + r_{\rm T}|\vec{\chi}|^2.
\end{align}
At a critical point where $r_{\rm E}=0$ but $r_{\rm T}>0$, fluctuations of $\vec{\chi}$ are suppressed by a mass $r_{\rm T}>0$, and vice versa. Consequently, a large difference $|r_{\rm E}-r_{\rm T}|$ can efficiently suppress two or three components of the fluctuating field. Since $\mbox{tr}(\phi^\dagger_{\text{E}}\phi_{\text{E}}\phi^\dagger_{\text{E}}\phi_{\text{E}})=\frac{4}{3}|\vec{\eta}|^4+\frac{2}{3}|\vec{\eta}^2|^2$ and $\mbox{tr}(\phi^\dagger_{\text{T}}\phi_{\text{T}}\phi^\dagger_{\text{T}}\phi_{\text{T}})=2|\vec{\chi}|^4$, the quartic term $Q_3$ can be expressed in terms of $Q_1$ and $Q_2$ in the cubic case. For the three-component case, however, a further cubic symmetric quartic term $Q_{\rm C}=\sum_{a<b}|\chi_a|^2|\chi_b|^2$ is allowed. Consequently, after a proper redefinition of the couplings, critical fluctuations of $\phi_{\rm E}$ and $\phi_{\rm T}$ are described by the effective actions
\begin{align}
 \label{cub4} S_{\rm E} = \int\mbox{d}^3x\Bigl[ \eta_a^*(-\nabla^2\delta_{ab})\eta_b + \bar{\lambda}_1|\vec{\eta}|^4+\bar{\lambda}_2|\vec{\eta}^2|^2\Bigr]
\end{align}
and
\begin{align}
  \nonumber S_{\rm T} = \int\mbox{d}^3x\Bigl[{}& \chi_a^*(-\nabla^2\delta_{ab})\chi_b +\bar{\lambda}_1|\vec{\chi}|^4+\bar{\lambda}_2|\vec{\chi}^2|^2\\
 \label{cub5}&+\bar{\lambda}_{\rm C}\sum_{a<b}|\chi_a|^2|\chi_b|^2\Bigr],
\end{align}
respectively. Although, in principle, cubic symmetry allows for additional kinetic terms similar to the $K$-dependent contribution in Eq. (\ref{field12}), we omit them here as they likely are irrelevant at the fixed points of the theory. The actions $S_{\rm E}$ and $S_{\rm T}$ for $\lambda_{\rm C}=0$ reduce to the $N$-component model $S'$ studied in section \ref{SecFlowN} with $N=2$ and $N=3$, respectively.

\section{Renormalization group}\label{SecRG}

The effective action $S$ is defined with respect to an ultraviolet momentum cutoff scale $\Omega$ so that fluctuations of the order parameter are restricted to momenta $p\leq \Omega$. Fluctuations of the order parameter are incorporated by successively reducing $\Omega$ which results in a running of the parameters of $S$. If the bosonic theory is derived from an underlying microscopic electronic model, the ultraviolet cutoff $\Omega$ corresponds to the infrared cutoff of the electron system. Typically, this infrared scale is given by temperature $T$ and we have $\Omega \sim T^{1/2}$. (We use nonrelativistic units $\hbar=k_{\rm B}=2M=1$ with electron mass $M$.) Note that the electron system itself has an ultraviolet momentum cutoff $\Lambda$ delimiting its applicability, often related to the bandwidth or Debye frequency.

In the following we discuss the RG flow for the running of couplings of the critical action $S$ (with $r=0$) as the ultraviolet momentum cutoff of the theory is changed from $\Omega$ to $\Omega/b$ with $b>1$ \cite{herbutbook}. In order to address the phase structure of the theory in three dimensions, we define several deformations of the theory that can be treated perturbatively close to the critical dimension of four by an expansion in $\vare=4-d$ with $0<\vare\ll1$. The expansion around four dimensions is complicated by the fact that $S$ intimately links the dimension of space to the number of field components and independent quartic couplings. Therefore, the RG evolution will be discussed for the following three limits:
\begin{itemize}
\item[A.] The weak-coupling limit of the three-dimensional model with $\text{SO}(3)\times \text{U}(1)$ symmetry, where the Gell-Mann algebra and angular momentum integrations are performed for $d=3$
\item[B.] The generalization of the model to $d$ dimensions with $\text{SO}(d)\times \text{U}(1)$ symmetry so that both Gell-Mann algebra and angular momentum integrations are consistently performed in $d$ dimensions, with a subsequent $\vare$-expansion
\item[C.] The extension of the $\text{SO}(5)\times \text{U}(1)$ symmetric  model with $K=0$ and $\lambda_3=0$ to an arbitrary number $N$ of field components $\vphi_a$
\end{itemize}
The RG flows of the three distinct cases are qualitatively compatible and thus give an idea of the critical behavior of the theory. In this section we focus on discussing the implications of the RG equations on the phase structure of the model, whereas the beta functions are derived in App. \ref{AppRG}.

To one-loop order, we only have to determine the fluctuation corrections to the three quartic couplings $\lambda_{1,2,3}$. Indeed, diagrammatically it is clear that at one-loop order there is no anomalous dimension $\eta$ or change of the parameter $K$. We can further assume that the system is fine-tuned to the transition such that $r=0$ after fluctuation contributions to $r$ have already been taken into account. To higher-loop order we expect $\eta>0$ and 
\begin{align}
 \frac{\mbox{d}K}{\mbox{d}\ln b} = \eta_K K
\end{align}
with $\eta_K<0$ so that $K$ becomes an irrelevant perturbation. This expectation derives from the behavior of corresponding couplings in similar models for real order in three-dimensional Heisenberg or tensor models \cite{herbutbook,PhysRevB.92.045117}. Therefore we can treat $K$ as a small perturbation, which, however, leads to some qualitative effects.

\subsection{Three-dimensional model}\label{SecFlow3D}

We first consider the three-dimensional action $S$ in Eq. (\ref{flow0}), where the indices of $\phi_{ij}$ take values $i,j=1,2,3$, the field has five complex components $\vphi_a$, and angular integrations are performed in three dimensions. For $K=0$ the one-loop running of the quartic couplings is given by
\begin{align}
 \nonumber \dot{\lambda}_1 = &{}\vare\lambda_1-18\lambda_1^2-8\lambda_1\lambda_2-8\lambda_2^2-\frac{160}{3}\lambda_1\lambda_3\\
 \label{flow1}&-32\lambda_2\lambda_3-\frac{272}{9}\lambda_3^2,\\
 \label{flow2} \dot{\lambda}_2 = &{}\vare\lambda_2-12\lambda_1\lambda_2-10\lambda_2^2-\frac{56}{3}\lambda_2\lambda_3+\frac{112}{9}\lambda_3^2,\\
 \label{flow3} \dot{\lambda}_3 = &{}\vare\lambda_3 -12\lambda_1\lambda_3+8\lambda_2\lambda_3-\frac{124}{3}\lambda_3^2.
\end{align}
We rescale the couplings according to $\lambda_m=\frac{1}{2\pi^2}\Omega^{-\vare}\bar{\lambda}_m$ and define $\dot{\lambda}_m=\mbox{d} \lambda_m/\mbox{d}\ln b$. We further introduce $\vare$ in the linear terms of the flow equations. By formally treating $0 < \vare \ll 1$ as a small parameter we can address the weak-coupling limit, because any fixed point of the set of the beta functions has coupling values $\lambda_{m\star} = \mathcal{O}(\vare)$.

The flow equations (\ref{flow1})-(\ref{flow3}) have a couple of remarkable properties that are related to the symmetries of the theory when $K=0$, which we discussed below Eq. (\ref{field19}): (i) First, for $\lambda_2=\lambda_3=0$ we have $\dot{\lambda}_2=\dot{\lambda}_3=0$. Indeed, if these two couplings vanish for some value of $\Omega$, the theory has an enlarged $\text{U}(5)$ symmetry. Consequently, the fluctuations of this theory cannot generate terms $|\vec{\vphi}^2|^2$ or $\mbox{tr}(\phi^\dagger\phi\phi^\dagger\phi)$ in the effective Lagrangian for $b>1$, as these terms do not possess this symmetry. (ii) Similarly, if $\lambda_3=0$, we have $\dot{\lambda}_3=0$, which can be ascribed to the $\text{SO}(5)\times \text{U}(1)$ symmetry of the theory with $\lambda_3=0$. This also implies that the coupling $\lambda_3$ cannot change sign during the RG flow, since this would require it to cross the plane of $\lambda_3=0$. (iii) Note also that even if $\lambda_2=0$ at some scale, the coupling is generated from a term in $\dot{\lambda}_2$, $\frac{112}{9}\lambda_3^2$. This manifests the difference of our model for symmetric traceless tensors from the field theory for the matrix model with $\text{U}(3)\times\text{U}(3)$ symmetry, where a nonzero $\lambda_2$ is forbidden by symmetry. Similarly, our flow equations for $\lambda_2=0$ do not reproduce those of the matrix model, see App. \ref{AppMat}.

The stability of a fixed point of the RG flow (defined by $\dot{\lambda}_m=0$ for all $m$) is determined by the number of relevant perturbations at the fixed point. We define the stability matrix at a fixed point $\lambda_{\star}$ by
\begin{align}
 \label{flow7} \mathcal{M}_{mn} = \frac{\partial \dot{\lambda}_m}{\partial \lambda_n}\Bigr|_{\lambda=\lambda_{\star}}.
\end{align} 
We denote the eigenvalues of $\mathcal{M}$ by $\{\theta_m\}$. A relevant (irrelevant) direction in the space of couplings is represented by a positive (negative) eigenvalue. Since temperature as the only tuning parameter is needed to fine-tune $r=0$, a second-order phase transition is described by a completely stable fixed point with all $\theta_m$ being negative. Although the flow of the parameters of the kinetic term $L_{\rm kin}$ vanishes at the one-loop level, it is fair to expect them to constitute irrelevant directions when including higher-loop orders.

\begin{figure}[t!]
\centering
\includegraphics[width=7.6cm]{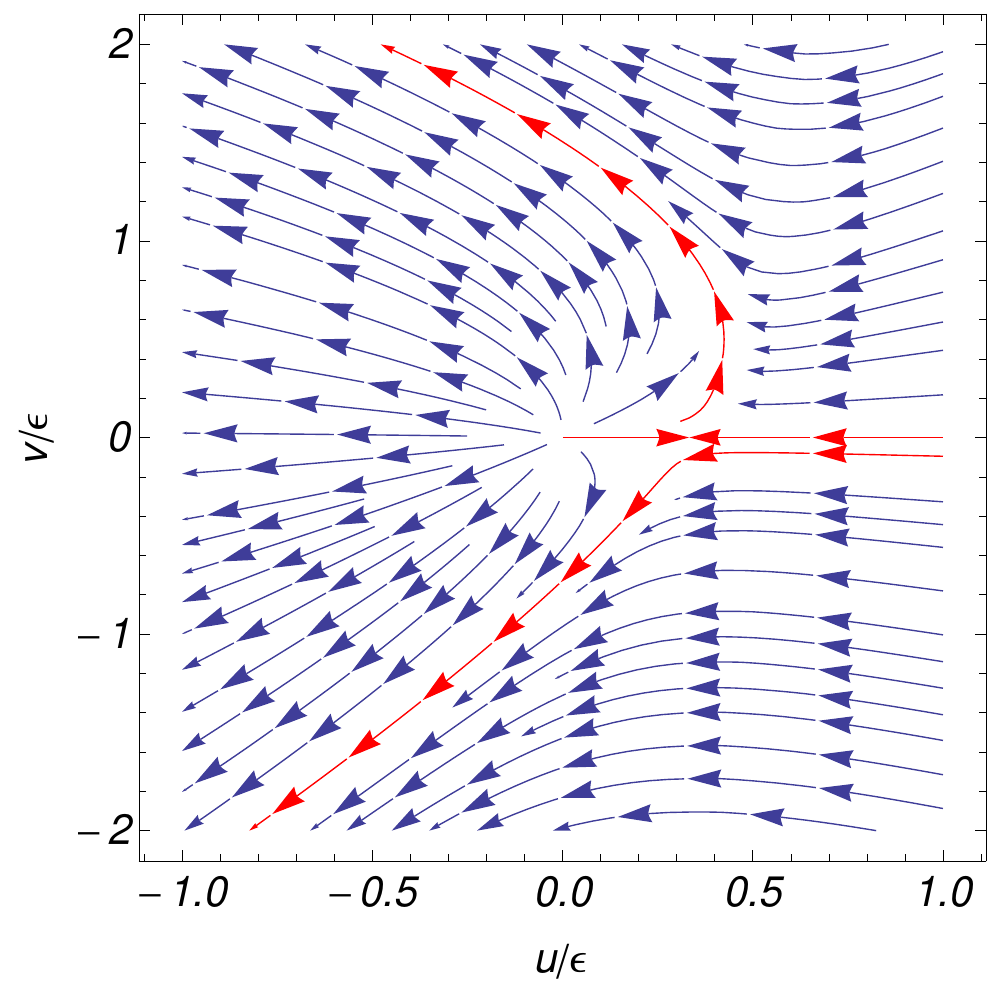}
\caption{Flow of the quartic couplings according to Eqs. (\ref{flow1})-(\ref{flow3}) in the plane spanned by $\lambda_3=0$. The flow equations in the plane $\lambda_3=0$ correspond to those of the $N$-component model discussed in Sec. \ref{SecFlowN} for $N=5$. We parametrize $\lambda_{1,2}$ in terms of $u=6(\lambda_1+\lambda_2)$ and $v=12\lambda_2$. This parametrization is useful because (i) the coupling $v$ cannot change sign and (ii) the quartic potential is stable if and only if $u>0$. The arrows of stream lines point towards the infrared, so that $b$ increases along the flow. We observe a runaway flow towards $u<0$ for all generic initial values of $\lambda_{1,2}$. This signals that fluctuations induce a first-order phase transition. Trajectories in the vicinity of the interacting fixed point $(u_\star,v_\star)=(\frac{\vare}{3},0)$ are shown in red.}
\label{FigFlowN5}
\end{figure}

The set of Eqs. (\ref{flow1})-(\ref{flow3}) features two fixed points. The non-interacting fixed point $(\lambda_1,\lambda_2,\lambda_3)=(0,0,0)$ is completely unstable with $(\theta_1,\theta_2,\theta_3)=(\vare,\vare,\vare)$. The interacting fixed point is located at
\begin{align}
 \label{flow8} (\lambda_{1\star}, \lambda_{2\star},\lambda_{3\star}) = \Bigl(\frac{\vare}{18},0,0\Bigr),
\end{align}
and thus possesses an emergent $\text{U}(5)$ symmetry. It is unstable in two directions with eigenvalues
\begin{align}
 \label{flow9} (\theta_1,\theta_2,\theta_3) =\Bigl(-\vare,\frac{\vare}{3},\frac{\vare}{3}\Bigr).
\end{align}
The flow of couplings is thus repelled from both fixed points, leading to a runaway flow in the space of couplings.

The flow in the representative plane spanned by $\lambda_3=0$ is shown in Fig. \ref{FigFlowN5}. As $b>1$ increases we first observe that $|\lambda_2|$ increases. This induces a significant running of $\lambda_1$, and eventually at $b=b_0>1$ the couplings flow into the unstable region with $\lambda_1+\lambda_2<0$. The negative sign of $\lambda_1+\lambda_2$ for $b>b_0$ signals a fluctuation-induced first-order phase transition: For $r=0$, the assumption of the effective action being of the form (\ref{flow0}) is inconsistent. Instead, a first-order transition occurs for a certain $r>0$, and higher orders in the field need to be included in the effective action to compute the minimum of the Ginzburg--Landau free energy.

For $K\neq 0$ the symmetry of the theory is reduced to $\text{SO}(3)\times \text{U}(1)$ and none of the quartic couplings is prohibited by an enlarged symmetry. The corresponding flow equations for arbitrary $K$ are displayed in App. \ref{AppRG}. Here we only discuss the flow equations for small $K$, as they exhibit many of the characteristic features of the full equations. We have
\begin{widetext}
\begin{align}
 \nonumber \dot{\lambda}_1 = &{}\vare \lambda_1 -\Bigl(18+\frac{692}{45}K^2\Bigr)\lambda_1^2-\Bigl(8+\frac{48}{5}K^2\Bigr)\lambda_1\lambda_2-\Bigl(8+\frac{208}{45}K^2\Bigr)\lambda_2^2 -\Bigl(\frac{160}{3}+\frac{7072}{135}K^2\Bigr)\lambda_1\lambda_3\\
 \label{flow4} &-\Bigl(32+\frac{3968}{135}K^2\Bigr)\lambda_2\lambda_3-\Bigl(\frac{272}{9}+\frac{14368}{405}K^2\Bigr)\lambda_3^2 +\mathcal{O}(K^3),\\
 \nonumber \dot{\lambda}_2 = &{}\vare\lambda_2 -\frac{4}{15}K^2\lambda_1^2-\Bigl(12+\frac{376}{45}K^2\Bigr)\lambda_1\lambda_2-\Bigl(10+\frac{148}{15}K^2\Bigr)\lambda_2^2+\frac{112}{135}K^2\lambda_1\lambda_3-\Bigl(\frac{56}{3}+\frac{592}{45}K^2\Bigr)\lambda_2\lambda_3\\
 \label{flow5} &+\Bigl(\frac{112}{9}+\frac{4688}{405}K^2\Bigr)\lambda_3^2 +\mathcal{O}(K^3),\\
 \nonumber \dot{\lambda}_3 =&{} \vare\lambda_3 -\frac{2}{15}K^2(\lambda_1^2-4\lambda_1\lambda_2+2\lambda_2^2)-\Bigl(12+\frac{128}{15}K^2\Bigr)\lambda_1\lambda_3+\Bigl(8+\frac{64}{9}K^2\Bigr)\lambda_2\lambda_3\\
 \label{flow6} &-\Bigl(\frac{124}{3}+\frac{4144}{135}K^2\Bigr)\lambda_3^2 +\mathcal{O}(K^3).
\end{align}
\end{widetext}
Clearly, for nonzero $K$ all couplings are generated, even if they are initially absent. Note that $K$ can have both positive or negative sign. However, since the first corrections are of order $K^2$, there is an accidental symmetry $K\to -K$ of the flow for small $K$. Expanding the full beta functions further, however, terms of order $K^3$ appear that break this invariance. The behavior for small $K$ is rooted in the fact that $D(\textbf{p})^{-1} = \frac{1}{p^2}\mathbb{1}+\mathcal{O}(K^2)$, and therefore no contribution linear in $K$ can be generated perturbatively, see App. \ref{AppRG}. For the same reason, terms linear in $K$ cannot appear to any perturbative loop order. If $\lambda_3$ is initially absent, a negative (positive) coupling $\lambda_3$ is generated if $\lambda_2/\lambda_1<\rho$ (if $\lambda_2/\lambda_1>\rho$) with $\rho=1-\frac{1}{\sqrt{2}}=0.293$. (We assume here $\lambda_2/\lambda_1\leq 1$ as it is relevant for the applications below.) We generically find that the explicit appearance of $K$ in the beta functions for $\lambda_{1,2}$ quantitatively modifies the flow of both couplings, whereas the effect of a nonzero value of $\lambda_3$ is small.

\begin{figure}[t!]
\centering
\includegraphics[width=7.6cm]{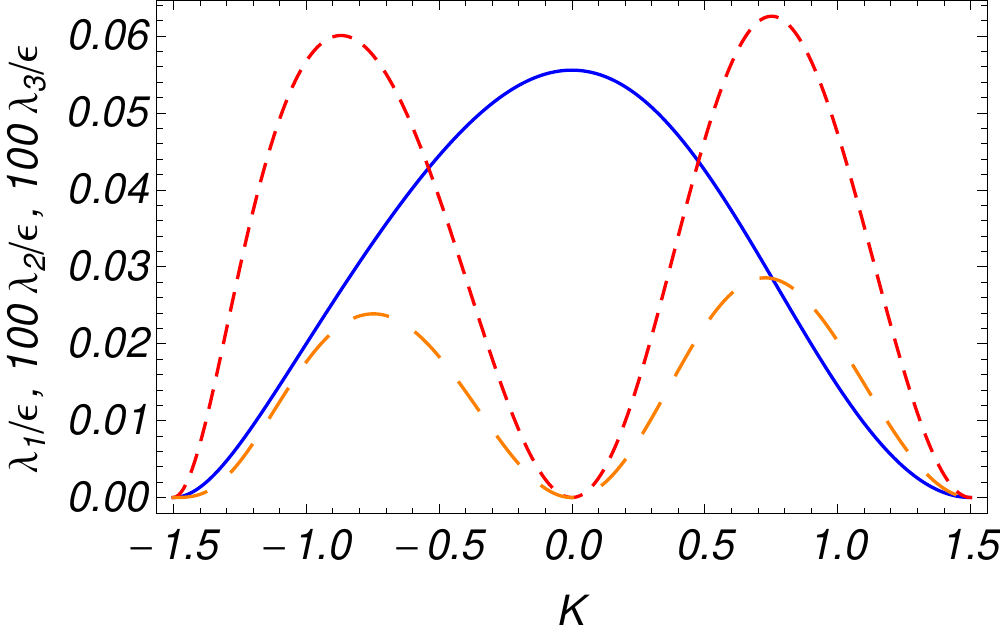}
\caption{Fixed point structure of the three-dimensional theory with action $S$ as a function of $K$. The plot shows the couplings $\lambda_{1\star}/\vare$ (blue, continuous line), $100\lambda_{2\star}/\vare$ (red, short-dashed line), and $100\lambda_{3\star}/\vare$ (orange, long-dashed line).  For better visibility, we magnify $\lambda_{2\star}$ and $\lambda_{3\star}$ by a factor of 100. To determine the couplings we use the full $K$-dependent expressions (\ref{rg30})-(\ref{rg32}), which yield Eqs. (\ref{flow4})-(\ref{flow6}) for small $K$. The eigenvalues of the stability matrix at the fixed point are $\theta_1=-\vare$ and positive $\theta_{2,3}=\mathcal{O}(\vare)$ for all values of $K$.}
\label{FigFixA}
\end{figure}

The fixed point structure is only mildly affected by the parameter $K\neq 0$. For small $K$, the interacting fixed point is shifted towards
\begin{align}
 \label{flow6b} \lambda_{1\star}&=\frac{1}{18}\Bigl(1-\frac{1142}{1215}K^2\Bigr)\vare+\mathcal{O}(K^3),\\
 \label{flow6c} \lambda_{2\star}&=\frac{K^2}{405}\vare+\mathcal{O}(K^3),\ \lambda_{3\star}=\frac{K^2}{810}\vare+\mathcal{O}(K^3).
\end{align}
As $K$ is increased further, $\lambda_{2\star}$ and $\lambda_{3\star}$ remain comparably small. However, in the extreme limit the fixed point couplings vanish: $\lambda_{m\star}\to 0$ as $|K|\to 3/2$. The stability of the fixed point is not changed qualitatively due to $K$, as it always features one irrelevant direction with $\theta_1=-\vare$, and two relevant directions with positive $\theta_{2,3}=\mathcal{O}(\vare)$. The behavior of $\{\lambda_m\}$ as a function of $K$ is displayed in Fig. \ref{FigFixA}.

\subsection{$d$-dimensional model}\label{SecFlow4D}

In this section we study the RG evolution of a complex tensor field close to its critical dimension of $d=4$. The corresponding action for an irreducible second-rank tensor under $\text{SO}(d)\times\text{U}(1)$ is given by
\begin{align}
 \nonumber S_d = {}& \int\mbox{d}^dx \Biggl[\vphi_a^*\Bigl(-\delta_{ab}\nabla^2+K\sqrt{\frac{d-1}{2d}}J_{abc}d_c(-\rmi \nabla)\Bigr)\vphi_b \\
 \label{eps1} &+ \bar{g}_1|\vec{\vphi}|^4+\bar{g}_2|\vec{\vphi}^2|^2+\bar{g}_3\mbox{tr}(\phi^\dagger\phi\phi^\dagger\phi)+\bar{g}_4\mbox{tr}(\phi^\dagger{}^2\phi^2)\Biggr].
\end{align}
We parametrize the tensor as $\phi_{ij} = \vphi_a\Lambda^a_{ij}$ with $i,j=1,\dots,d$, and $a,b,c=1,\dots,N_d$, where $N_d= \frac{(d-1)(d+2)}{2}$ is the number of real Gell-Mann matrices in $d$ dimensions \cite{PhysRevB.92.045117}. We define
\begin{align}
 \label{eps3} d_a(\textbf{p}) = \sqrt{\frac{d}{2(d-1)}}p_ip_j\Lambda^a_{ij}
\end{align}
so that $\sum_a d_a^2=p^4$. When computing perturbative corrections derived from $S_d$, the angular momentum integrals can be computed in $d$ dimensions. In the interaction part of $S_d$ we account for the fact that $\mbox{tr}(\phi^\dagger{}^2\phi^2)$ is not a function of the remaining three quartic terms when $d>3$. For this reason, we labelled the quartic couplings by $\{\bar{g}_m\}$ to distinguish them from $\{\bar{\lambda}_m\}$.

The action $S_{d=4}$ consistently lifts all the algebraic and analytic properties of the three-dimensional model to four dimensions. In particular, for the number of field components $\vphi_a$ to be consistent with a second-rank tensor in four-dimensional space, we need to replace the $N_3=5$ components of the original model by $N_4=9$ components. The nine Gell-Mann matrices in four dimensions are given in Eq. (\ref{gell1}). The extension from five to nine functions $d_a(\textbf{p})$ was first employed by Abrikosov in order to perform an $\vare$-expansion for the three-dimensional fermionic QBT system \cite{abrikosov}.

The RG evolution of the quartic couplings for small $\vare$ and to leading order in $K$ are given in Eqs. (\ref{eps4})-(\ref{eps7}) in the appendix, where we rescale the couplings according to $g_m=\frac{1}{8\pi^2}\bar{g}_m$. The leading order in $K$ is again found to be $\mathcal{O}(K^2)$. In four dimensions, stability of the kinetic operator requires $-1 < K < 2$. Analogous to the three-dimensional model, for $K=0$, the couplings $g_3$ and $g_4$ cannot be generated from $g_1$ and $g_2$ if initially absent, because the theory has an enlarged $\text{SO}(N_d)\times \text{U}(1)$ symmetry for $g_3=g_4=0$. For $K\neq 0$, however, both couplings are generated even if initially absent.

The fixed point structure of the flow equations is analogous to the behaviour of the three-dimensional model $S$ discussed in the previous section. We have the non-interacting fixed point $(g_1,g_2,g_3,g_4)=(0,0,0,0)$, which is completely unstable with $(\theta_1,\theta_2,\theta_3,\theta_4)=(\vare,\vare,\vare,\vare)$. An interacting fixed point with enlarged $\text{U}(9)$ symmetry is found for
\begin{align}
 \label{eps8} (g_{1\star},g_{2\star},g_{3\star},g_{4\star}) = \Bigl( \frac{\vare}{26},0,0,0\Bigr),
\end{align}
with stability eigenvalues
\begin{align}
 \label{eps9} (\theta_1,\theta_2,\theta_3,\theta_4) = \Bigl(-\vare,\ \frac{7}{13}\vare,\ \frac{7}{13}\vare,\ \frac{7}{13}\vare\Bigr).
\end{align}
It thus constitutes a four-dimensional analogue of the interacting fixed point of the three-dimensional model in Eq. (\ref{flow8}). Hence, both the structure of the beta-functions and the fixed points of the RG flow obtained here are similar to those obtained in Sec. \ref{SecFlow3D}. This supports the results obtained for the three-dimensional model and the conclusions drawn from them. Since the $d$-dimensional model with $d>3$ has four quartic terms instead of three, it is difficult to directly physically interpret the running of couplings for $d>3$ in the context of superconductors.

\subsection{$N$-component model}\label{SecFlowN}

From the findings of the previous sections we conclude that in order to make the critical properties of $S$ particular transparent, it is advantageous to focus on the plane spanned by $(\lambda_1,\lambda_2,\lambda_3,K)=(\lambda_1,\lambda_2,0,0)$ in coupling space. RG trajectories within this plane are protected by symmetry to stay within this plane. We generalize the setup to $N$ field-components $\vphi_a$ by considering the action
\begin{align}
\label{Ncomp1} S' = \int\mbox{d}^d x\Bigl[ \vphi^*_a(-\nabla^2)\vphi_a+\bar{\lambda}_1|\vec{\vphi}|^4+\bar{\lambda}_2|\vec{\vphi}^2|^2\Bigr]
\end{align}
with $a=1,\dots,N$. The system described by $S'$ is invariant under $\text{SO}(N)\times\text{U}(1)$ transformations. 

We study the running of the couplings $\lambda_{1,2}$ of $S'$ by means of an $\vare$-expansion around $d=4$ dimensions. After rescaling of the couplings according to $\lambda_m=\frac{1}{8\pi^2}\bar{\lambda}_m$ we obtain the flow equations
\begin{align}
 \label{Ncomp2} \dot{\lambda}_1 &= \vare \lambda_1 -2(N+4)\lambda_1^2-8\lambda_1\lambda_2-8\lambda_2^2,\\
 \label{Ncomp3} \dot{\lambda}_2 &=\vare \lambda_2 -12\lambda_1\lambda_2-2N \lambda_2^2.
\end{align}
For $N=5$ we recover the flow equations of Sec \ref{SecFlow3D} for $K=0$ and $\lambda_3=0$. Similarly, the flow equations of Sec. \ref{SecFlow4D} for $K=0$ and $(g_1,g_2,g_3,g_4)=(\lambda_1,\lambda_2,0,0)$  are obtained for $N=9$.

The set of Eqs. (\ref{Ncomp2}) and (\ref{Ncomp3}) has the following fixed point structure as $N$ is varied. The non-interacting fixed point $(\lambda_1,\lambda_2)=(0,0)$ is repulsive with $(\theta_1,\theta_2)=(\vare,\vare)$. For all $N$ there exists an interacting ``Heisenberg'' fixed point ``H'' with $\text{U}(N)$ symmetry at
\begin{align}
 \label{Ncomp4}\text{H}:\  (\lambda_{1\star},\lambda_{2\star}) = \Bigl(\frac{\vare}{2(N+4)},0\Bigr),
\end{align}
with
\begin{align}
 \label{Ncomp5} (\theta_1,\theta_2) = \Bigl(-\vare,\frac{N-2}{N+4}\vare\Bigr).
\end{align}
We recognize this fixed point to be the generalization towards $N$ field components of the interacting fixed points found in Sec. \ref{SecFlow3D} ($N=5$) and Sec. \ref{SecFlow4D} ($N=9$). 

The fixed point H is stable for $N<2$ and unstable for $N>2$. The change in stability at $N=2$ can be understood from a stability exchange with yet another fixed point of the system. Indeed, the flow equations feature two more fixed points B$_1$ and B$_2$ with couplings
\begin{align}
 \label{Ncomp6} \text{B}_1:\ &(\lambda_{1\star},\lambda_{2\star}) = \Bigl(\frac{f-N\xi}{4h},\ \frac{g+3\xi}{2h}\Bigr )\vare,\\
\label{Ncomp7} \text{B}_2:\ &(\lambda_{1\star},\lambda_{2\star})= \Bigl(\frac{f+N\xi}{4h},\ \frac{g-3\xi}{2h}\Bigr)\vare,
\end{align}
where $f=48+N(N-4)$, $g=N(N+1)-12$, $\xi=\sqrt{48-24N+N^2}$, $h= 3f+Ng>0$. The fixed points $B_{1,2}$ are real for $48-24N+N^2>0$, which requires either $N<4(3-\sqrt{6})=2.20$ or $N>4(3+\sqrt{6})=21.8$ components of the field. As these critical values of $N$ are approached from below or above, respectively, the fixed points B$_1$ and B$_2$ collide ($\xi=0$) and mutually annihilate each other. For $N=2$, fixed point B$_1$ coincides with H ($\lambda_{1\star}=\frac{\vare}{12}$, $\lambda_{2\star}=0$) and exchanges stability with it. We have the following stable fixed points as $N$ varies:
\begin{itemize}
 \item[1)] For $N<N_{\rm H}$ four fixed points exist, only H is stable.
 \item[2)] For $N_{\rm H} < N <N_-$ four fixed points exist, only B$_1$ is stable with $\lambda_{2\star}<0$.
 \item[3)] For $N_- < N < N_+$, only the Gaussian and Heisenberg fixed points exist, and none of them is stable.
 \item[4)] For $N_+<N$, four fixed points exist, only B$_2$ is stable with $\lambda_{2\star}>0$.
\end{itemize}
To one loop order we have $N_{\rm H}=2+\mathcal{O}(\vare)$ and $N_{\pm}=4(3\pm\sqrt{6})+\mathcal{O}(\vare)$. The correlation length exponent $\nu$ is given by
\begin{align}
 \label{Ncomp10} \nu = \frac{1}{2}+\frac{N+1}{2}\lambda_{1\star} + \frac{1}{2} \lambda_{2\star} +\mathcal{O}(\vare^2)
\end{align}
so that for H we obtain $\nu=\frac{1}{2}+\frac{N+1}{4(N+4)}\vare+\mathcal{O}(\vare^2)$, which is the exponent at the Wilson--Fisher fixed point of the $\text{O}(2N)$ model, whereas the fixed points B$_1$ and B$_2$ lie within a distinct ``chiral'' universality class, as pointed out first by Kawamura \cite{Kawamura1,PhysRevB.38.4916}. The anomalous dimension for $N=2$ at both H and B$_1$ reads $\eta = \frac{\vare^2}{48} +\mathcal{O}(\vare^3)$ \cite{2001NuPhB.607..605P}, and the susceptibility exponent at H is related to $\nu$ and $\eta$ via the scaling relation $\gamma=\nu(2-\eta) = 2\nu +\mathcal{O}(\vare^2)$.

The coincidence of the fixed points H and B$_1$ for $N=2$ and their instability for $N\geq 3$ indicates that higher-loop terms in the RG beta functions play a crucial role for the cases $N=2,3$ that are relevant for superconductors with cubic anisotropy. The fixed point structure of $S'$ has been investigated in the context of frustrated magnetism. To see the correspondence write $\vphi_a = \frac{1}{\sqrt{2}}(\sigma_a + \rmi \tau_a)$, so that
\begin{align}
 \label{Ncomp11}S' ={}& \int\mbox{d}^dx \Bigl( \frac{1}{2}\sigma_a(-\nabla^2)\sigma_a+\frac{1}{2}\tau_a(-\nabla^2)\tau_a \\
  \nonumber &+ \frac{(\bar{\lambda}_1+\bar{\lambda}_2)}{4}(\vec{\sigma}^2+\vec{\tau}^2)^2+ \bar{\lambda}_2[(\vec{\sigma}\cdot\vec{\tau})^2-\vec{\sigma}^2\vec{\tau}^2]\Bigr),
\end{align}
which coincides with the spin model given by
\begin{align}
 \nonumber S' = \int\mbox{d}^dx\Bigl({}& \frac{1}{2}\sum_{\alpha} \vec{\psi}_{\alpha}\cdot(-\nabla^2)\vec{\psi}_\alpha+\frac{\bar{u}}{4!}\sum_\alpha (\vec{\psi}_\alpha^2)^2\\
 \label{Ncomp12} &+\frac{\bar{v}}{4!} \sum_{\alpha,\beta}((\vec{\psi}_\alpha\cdot\vec{\psi}_\beta)^2-\vec{\psi}_\alpha^2\vec{\psi}_\beta^2)\Bigr)
\end{align}
upon identifying $\vec{\psi}_1=\vec{\sigma}$,\ $\vec{\psi}_2=\vec{\tau}$ with $\alpha,\beta=1,2$, and $\bar{u}=6(\bar{\lambda}_1+\bar{\lambda}_2)$, $\bar{v}=12\bar{\lambda}_2$.

The spin model in Eq. (\ref{Ncomp12}) constitutes an $\text{O}(N)\times \text{O}(2)$ symmetric field theory that has been approached from various theoretical and experimental directions. The three-loop $\vare$-expansion yields \cite{1995PhLA..208..161A,PelissettoVicari}
\begin{align}
 \label{Ncomp13} N_{\rm H} &= 2 -\vare +1.29\vare^2 +\mathcal{O}(\vare^3),\\
 \label{Ncomp14} N_- &= 2.20 - 0.57\vare +0.99\vare^2 +\mathcal{O}(\vare^3),\\
 \label{Ncomp15} N_+ &= 21.80 - 23.43 \vare +7.09\vare^2 +\mathcal{O}(\vare^3).
\end{align}
We observe rather bad convergence properties to this loop order. The six-loop perturbative RG in fixed dimension $d=3$ predicts stable fixed points for both $N=2,3$ with $u_{\star},v_{\star}>0$ \cite{PhysRevB.63.140414}, although this is in conflict with three-loop order results in the same scheme. Note that the fixed points in fixed dimension $d=3$ need not be smoothly connected to H and $B_{1,2}$ that are described by Eqs. (\ref{Ncomp13})-(\ref{Ncomp14}). The predicted fixed points for $N=2,3$ correspond to second-order phase transitions with universal exponents, and since they feature $\lambda_{2\star}\neq 0$, they are in the chiral universality class. As is carefully discussed in Refs. \cite{PhysRevB.69.134413}, there exists strong evidence conflicting a second-order phase transition, crucially, a negative anomalous dimensions in Monte Carlo results, and distinct experimental systems that observe some scaling, but with non-universal exponents. Based on these observations and a study with the nonperturbative RG it is then argued that the transition is first-order but with an extended regime of scaling because of the proximity of a complex fixed point close to the RG trajectory.

For the cubic anisotropic system with $N=3$ the additional quartic term $\lambda_{\rm C}\sum_{a<b}|\chi_a|^2|\chi_b|^2 = \frac{\lambda_{\rm C}}{2}(|\vec{\chi}|^4-\sum_a |\chi_a|^4)$ introduces a further marginal operator that influences the RG flow. In particular, for $\lambda_2=0$ the system may feature a stable fixed point with cubic symmetry \cite{PelissettoVicari}. For the applications that are relevant here, however, we have $\lambda_2\neq 0$. For a discussion of the three-loop $\vare$-expansion of the model with $\lambda_{1,2,\rm C}$ see Ref. \cite{PhysRevB.49.15901}. The nonperturbative RG flow of the $\text{O}(3)\times \text{O}(2)$ symmetric model with $\lambda_{\rm C}=0$ and its relevance for experiments with spin-1 Bose--Einstein condensates has been laid out in Refs. \cite{PhysRevA.93.051603,PhysRevA.94.053623}.

\section{Superconductivity in Luttinger semimetals}\label{SecQBT}

In this section we discuss the significance of the structure of the RG flow for the complex tensor order superconducting transition in Luttinger semimetals. For this purpose we first derive the initial conditions for the bosonic RG flow from the underlying fermionic model at the mean-field level. We then discuss the implication of the running of couplings for the phase transitions of the system. Then we investigate the influence of cubic anisotropy.

The Lagrangian for spin-$3/2$ Luttinger electrons at an isotropic three-dimensional quadratic band touching point with local short-range interactions is given by
\begin{align}
 \nonumber L_{\psi} ={}& \psi^\dagger(\partial_\tau+d_a(-\rmi \nabla)\gamma_a-\mu)\psi+ g_{\rm s}(\psi^\dagger\gamma_{45}\psi^*)(\psi^{\rm T}\gamma_{45}\psi) \\
 \label{qbt2} &+g_{\rm d} (\psi^\dagger\gamma_a\gamma_{45}\psi^*)(\psi^{\rm T}\gamma_{45}\gamma_a\psi),
\end{align}
with $\psi$ a four-component Grassmann spinor,  $\{\gamma_a\}$ a set of five anti-commuting $4\times 4$ Dirac matrices such that $\gamma_{1,2,3}$ are real and $\gamma_{4,5}$ are imaginary, $\gamma_{45}=\rmi\gamma_4\gamma_5$, and coupling constants $g_{\rm s}$ and $g_{\rm d}$. The chemical potential is denoted by $\mu$. We use units such that $\hbar=k_{\rm B}=2M=1$ with  electron mass $M$. The Lagrangian for the electrons is defined with respect to an ultraviolet momentum cutoff $\Lambda$. A transition towards complex tensor order is induced for $g_{\rm s}=0$ and $g_{\rm d}=-g<0$.  At nonzero temperature $T>0$ we can integrate out the fermions on the mean-field level to obtain a Ginzburg--Landau theory for the order parameter field $\Delta_a=\langle \psi^{\rm T}\gamma_{45}\gamma_a\psi\rangle$.

When integrating out the fermions, we generate kinetic terms for the field $\Delta_a(\tau,\textbf{x})$, which result in fluctuations of the order parameter field close to the second-order phase transition with $\langle\Delta_a\rangle=0$. The corresponding Ginzburg--Landau effective action in the isotropic case is given by
\begin{align}
\nonumber S_{\Omega} ={}& \int_0^{1/T}\mbox{d}\tau \int\mbox{d}^3x \Biggl( \Delta_a^*\Bigl[r\delta_{ab}+Z \delta_{ab}\partial_\tau-X \delta_{ab}\nabla^2\\
 \label{qbt3} &+\frac{Y}{\sqrt{3}}J_{abc}d_c(-\rmi \nabla)\Bigr]\Delta_b+q_1 |\vec{\Delta}|^4+q_2|\vec{\Delta}^2|^2\Biggr).
\end{align}
This bosonic theory is defined with respect to an ultraviolet momentum cutoff $\Omega\sim T^{1/2}$ that we identify with the temperature of the microscopic fermionic theory. The coefficients $r,X,Y,Z,q_1,q_2$ are functions of $g, \mu, T,\Lambda$ and are given in App. \ref{AppIni}. Importantly, no quartic term $Q_3$ is generated at the mean-field level so that the quartic theory has an accidental $\text{SO}(5)\times\text{U}(1)$ symmetry. Since temporal fluctuations with frequency $\omega_n =2\pi nT$ ($n\in\mathbb{Z}$) are suppressed for $n\neq0$, the fluctuations close to the transition are $\tau$-independent: $\Delta_a(\tau,\textbf{x}) = \sum_n\Delta_a(\textbf{x},n)e^{\rmi \omega_n \tau}\equiv \Delta_a(\textbf{x})$. The critical theory with cutoff $\Omega$ can then be written as
\begin{align}
 \nonumber S_\Omega = \int\mbox{d}^3x\Biggl({}&\vphi_a^*\Bigl[-\delta_{ab}\nabla^2+\frac{K_\Omega}{\sqrt{3}}J_{abc}d_c(-\rmi \nabla)\Bigr]\vphi_b \\
 \label{qbt4}&+ \bar{\lambda}_{1\Omega}|\vec{\vphi}|^4+\bar{\lambda}_{2\Omega}|\vec{\vphi}^2|^2\Biggr),
\end{align}
where we applied a field redefinition $\vphi_a \propto T^{-3/4}\Delta_a$. From this expression for $S_\Omega$ the initial values of the couplings $K(b)$ and $\lambda_{1,2,3}(b)$ for $b=1$ can be read off. We explicitly construct Eq. (\ref{qbt4}) from Eq. (\ref{qbt3}) in App. \ref{AppIni}, where we also give the expressions for the initial values used in the following.

At the mean-field level, the fermion system with $g_{\rm s}=0$ and $g_{\rm d}=-g<0$ features two qualitatively distinct second-order phase transitions towards complex tensor order, which we refer to as the strong-coupling and weak-coupling transitions, respectively. The mean-field phase diagram is discussed in Ref. \cite{BoettcherComplex}. The strong coupling transition appears for couplings succeeding a critical value, $g>g_{\rm c}$, and for high relative temperatures $T_{\rm c}\sim \Lambda^2$. In particular, it persists for $\mu=0$. It features $\lambda_{2\Omega}<0$ and thus real nematic order develops at the transition. The weak-coupling transition, on the other hand, only appears for finite chemical potential $\mu>0$, at exponentially small critical temperatures $T_{\rm c}\ll \mu$, and for arbitrarily weak coupling $g$. Due to $\lambda_{2\Omega}>0$ at this transition, genuinely complex orders that break time-reversal symmetry are energetically favoured. We thus conclude that the strong- and weak-coupling transitions are initialized, respectively, in the lower and upper half-planes of the flow diagram for $u=6(\lambda_1+\lambda_2)$ and $v=12\lambda_2$ shown in Fig. \ref{FigFlowN5}.

\begin{figure}[t!]
\centering
\begin{minipage}{0.46\textwidth}
\includegraphics[width=7.6cm]{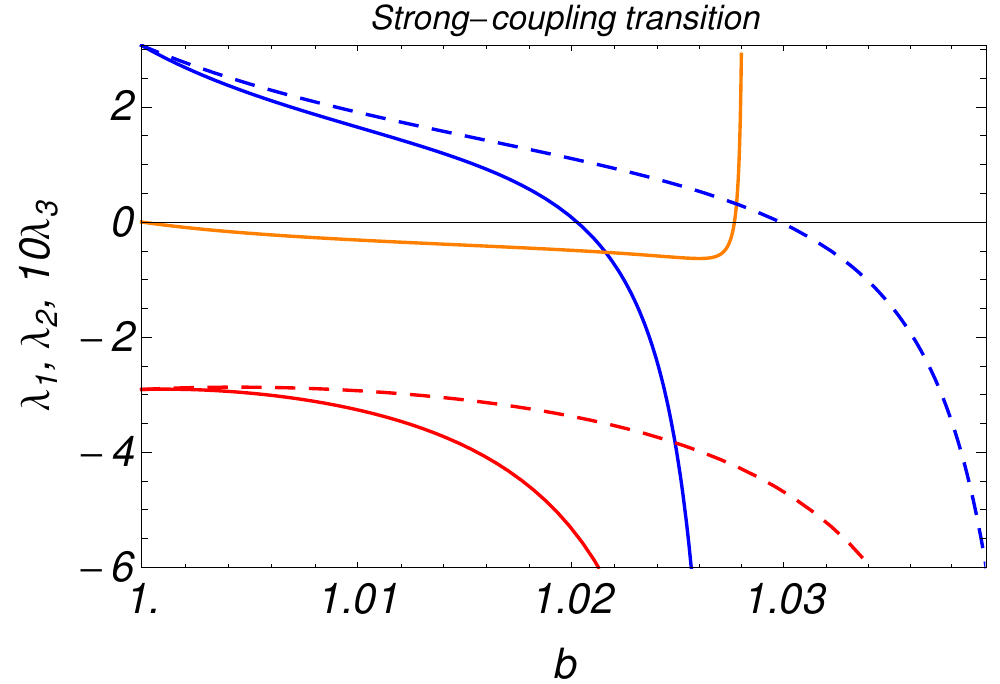}
\includegraphics[width=7.6cm]{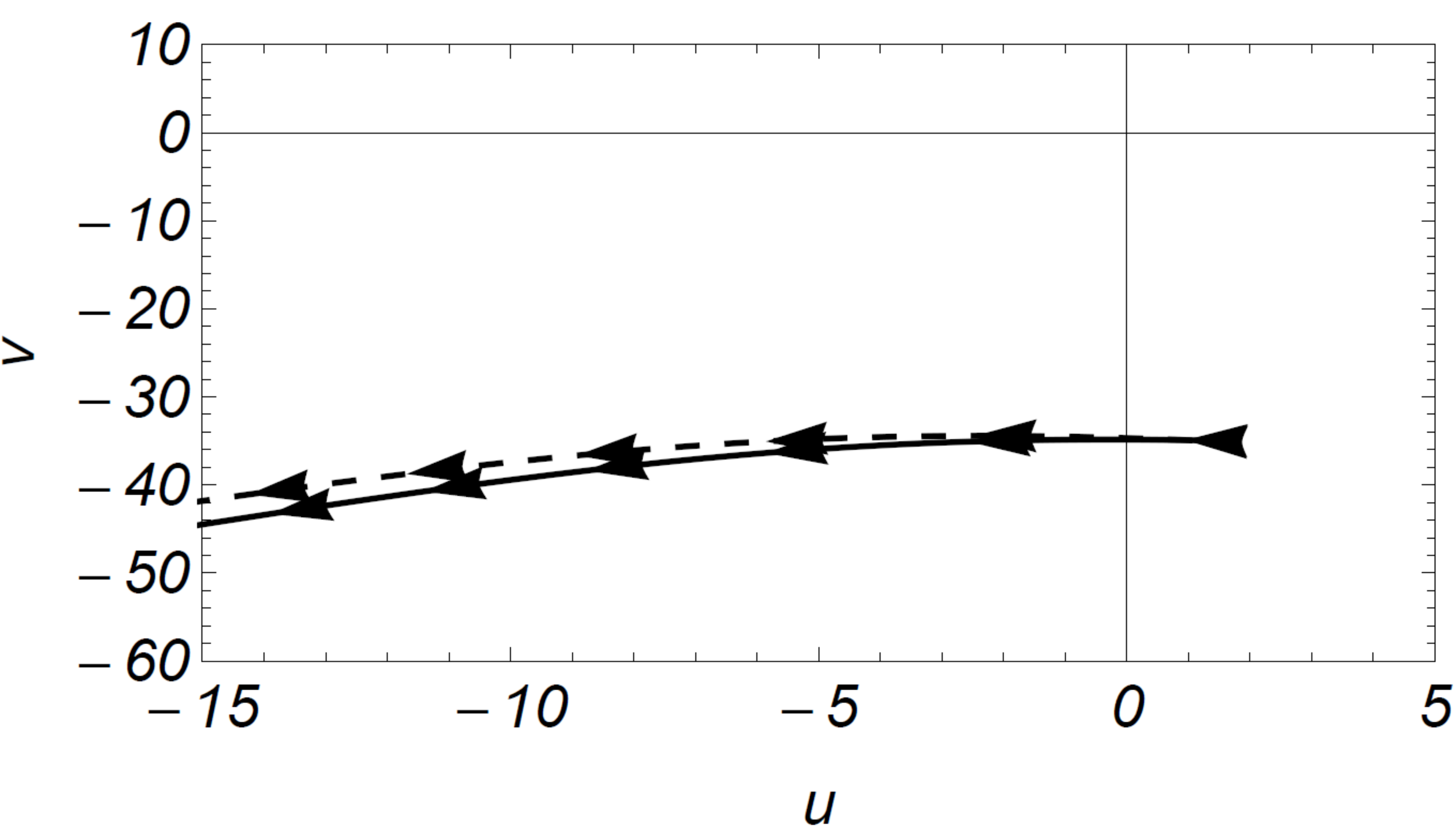}
\caption{Strong-coupling transition in isotropic Luttinger semimetals. \underline{Upper panel:} The solid lines, from top to bottom, show the scale evolution of the running couplings $\lambda_1(b)$ (blue line), $\lambda_3(b)$ (magnified by a factor of ten, orange line), $\lambda_2(b)$ (red line). They rapidly enter the region of $(-,-,+)$ in Table \ref{TabStab} and the stability criterion is violated because $\lambda_1+\lambda_2+\frac{4}{3}\lambda_3<0$ for $b>b_0\sim 1$, indicating a fluctuation-induced first-order transition. The dashed lines show the result of integrating the flow for $K=0$ (same colour scheme), which constitutes a reasonable approximation. \underline{Lower panel:} The solid line represents the evolution of the couplings $u=6(\lambda_1+\lambda_2)$ and $v=12\lambda_2$ projected onto the plane with $\lambda_3=0$, see Fig. \ref{FigFlowN5} for comparison. Arrows point towards the infrared. The dashed line again constitutes the flow when setting $K=0$. In the latter case, the instability is indicated by $u<0$.
\\ In the plots we chose $T/\Lambda^2=0.3$ as the sole input parameter and set $\vare=1$.}
\label{FigStrong}
\end{minipage}
\end{figure}

We first study the strong-coupling transition for $\mu=0$ with critical temperatures $T_{\rm c}/\Lambda^2>0.291$. (For smaller $T_{\rm c}/\Lambda^2$ the mean-field transition is of first order.) In this regime we find $\lambda_{1\Omega}\sim 1$ and $\lambda_{2\Omega}\sim -1$ so that $u_\Omega=6(\lambda_{1\Omega}+\lambda_{2\Omega})\gtrsim0$ and $K_\Omega \sim 0.5$. Consequently, a negative coupling $\lambda_3$ is generated during the flow. The effect of the nonzero $\lambda_3$ onto the running of $\lambda_{1,2}$ is negligible. In particular, due to the large value of $|\lambda_{2\Omega}|$ the flow trajectories rapidly enter the instability region of $u(b)<0$ for $b>b_0$ with $b_0\sim 1$. As a result, a first-order transition is induced due to fluctuations. The flow is depicted in Fig. \ref{FigStrong}.

\begin{figure}[t!]
\centering
\begin{minipage}{0.46\textwidth}
\includegraphics[width=7.6cm]{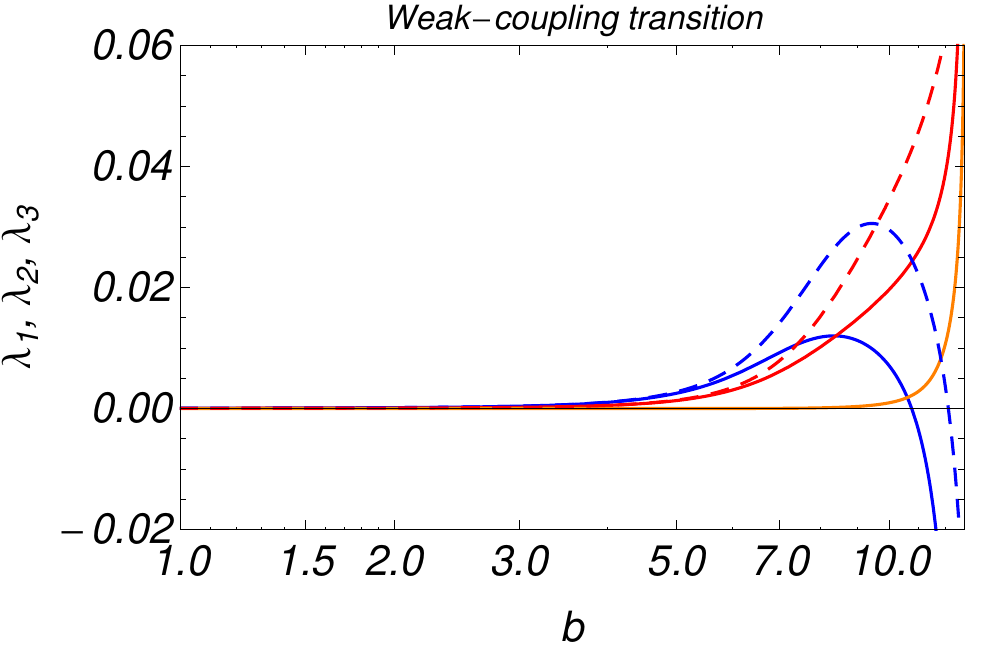}
\includegraphics[width=7.6cm]{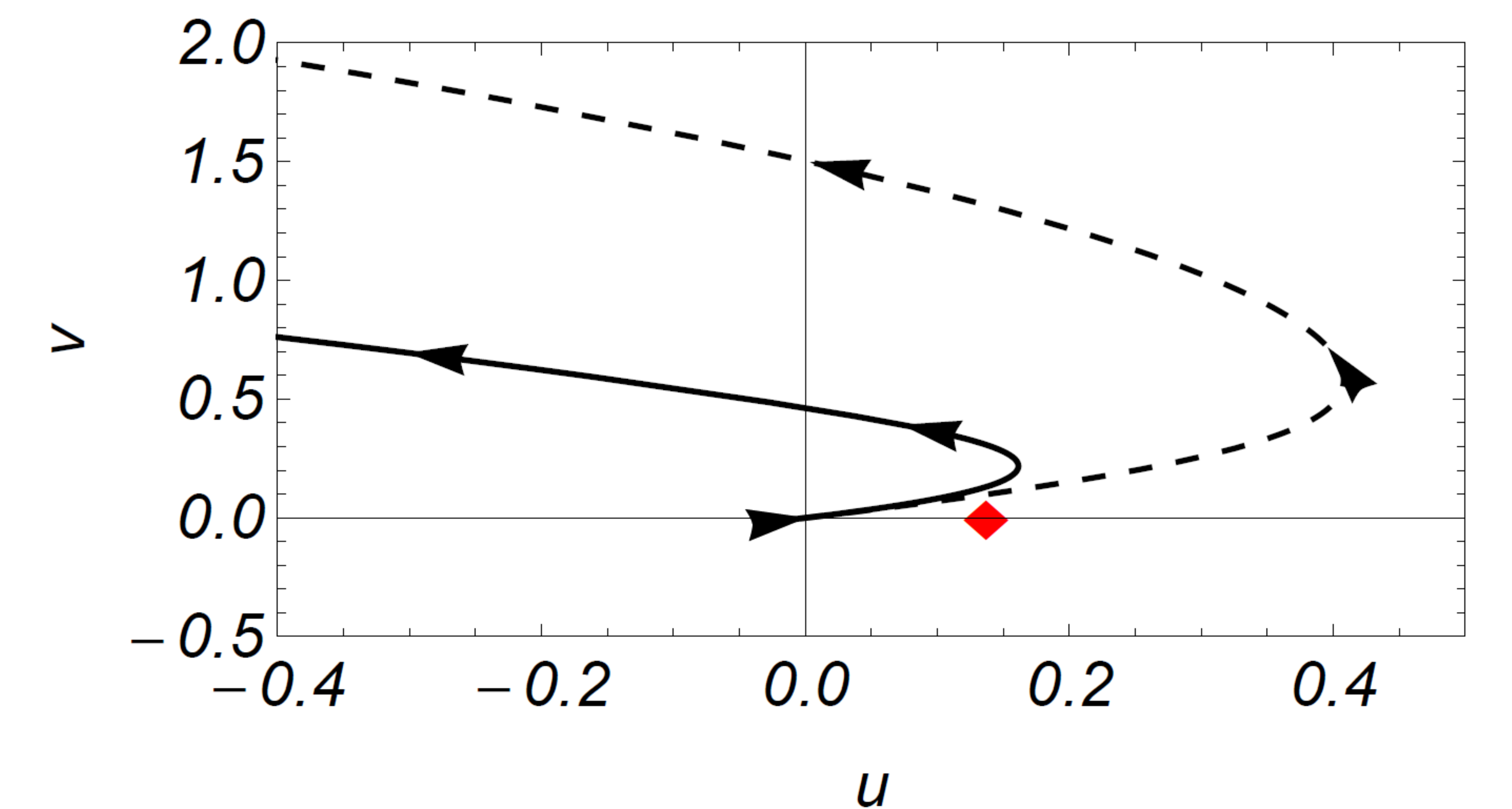}
\caption{Weak-coupling transition in isotropic Luttinger semimetals. The colour and line scheme is chosen as in Fig. \ref{FigStrong} for the strong-coupling transition. Therefore, we focus on pointing out the differences here. \underline{Upper panel:} The flow enters the regime of $(-,+,+)$ in Table \ref{TabStab} after a relatively long RG-time $b$. An instability is signalled by $\lambda_1+\frac{4}{3}\lambda_3<0$ for $b>b_0\sim 10$. Due to the large value of $b_0$, the fluctuation-induced first-order transition is only weak. \underline{Lower panel:} The slowing down of the flow before entering the region of instability can be explained by the flow trajectory being close to the interacting fixed point (red mark). This proximity, in contrast to the flow in Fig. \ref{FigStrong}, is induced by the small initial couplings. The dashed line representing the flow for $K=0$ deviates more significantly since the interacting fixed point for $K=0$ is further to the right in the plot, see Fig. \ref{FigFixA}.
\\ In the plots we choose $\mu/\Lambda^2=0.4$ and $T/\mu=0.005$, which are values motivated from half-Heusler superconductors \cite{BoettcherComplex}, and set $\vare=1$.}
\label{FigWeak}
\end{minipage}
\end{figure}

At the weak-coupling transition for $g\to 0$ and $T_{\rm c}/\mu\ll 1$ we have $\lambda_{1\Omega}=2\lambda_{2\Omega}>0$. In this regime we have $K_\Omega=0.86$ and due to the ratio $\lambda_{2\Omega}/\lambda_{1\Omega}=\frac{1}{2}$ being larger than $1-\frac{1}{\sqrt{2}}=0.29$ a positive coupling $\lambda_3$ is induced. The initial values of $\lambda_{1,2}$ are tiny so that the flow is initialized close to the two fixed points and the flow of $\lambda_3$ has a quantitative but small influence on the other two couplings. The RG trajectories are repelled from the fixed points and eventually flow into a region of instability, again indicating a fluctuation-induced first-order transition. However, due to the small initial values, the running couplings $\lambda_m(b)$ stay in a region of parameter space with stable quartic potential for a sufficiently long RG time $b<b_0$ with $b_0\sim 10$. The slowing down of the flow in the stable region results from the vicinity of the interacting fixed point. Further, this proximity of the RG trajectory to the interacting fixed point might yield a certain range of temperatures where scaling is observed, despite the transition being of first order. The flow is visualized in Fig. \ref{FigWeak}.

The size of the scale $b_0$ at which the flow enters the region of instability gives an indication on the temperature range where critical fluctuations are important and the size of the jump of the order parameter. For $b_0\gg1$, as occurs for the weak-coupling transition with $b_0\sim 10$, only the temperature interval $|T-T_{\rm c}|\sim T_{\rm c}/b_0^2$ in the vicinity of the transition temperature shows the fluctuation-induced first-order transition. Outside this temperature range, mean-field theory gives a solid estimate of the size of the order parameter. The jump of the order parameter is thus delimited by the mean-field value for the gap $\Delta_{\rm MF}\sim |T-T_{\rm c}|$ to be $\Delta_{\rm MF}/b_0^2$. Consequently, both the critical region and the order parameter jump are very small at the weak-coupling transition and it may thus appear continuous in experiment or numerical simulation. If $b_0$ is not particularly large, as in the case of the strong-coupling transition, a similar estimate is more difficult to make. In fact, as we expound in the following section, the strong-coupling transition should for several reasons rather be treated with a method that incorporates both fermionic and bosonic fluctuations on an equal footing such as the Functional Renormalization Group.

The quartic coupling $\lambda_3\neq 0$ that is generated during the RG flow removes the accidental $\text{SO}(5)\times\text{U}(1)$ symmetry of the quartic mean-field free energy. Since the effect is small, however, we can treat $\lambda_3$ as a small correction such that the signs of the couplings $\lambda_{1,2}$ determine whether the order parameter is real ($|\vec{\Delta}^2|=|\vec{\Delta}|^2$) or genuinely complex ($\vec{\Delta}^2=0$), and the sign of $\lambda_3$ determines the energy of distinct states within this manifold of potential ground states. For the strong coupling transition, we have $\lambda_1>0$ and $\lambda_2<0$ and real orders are favoured, so that the additional quartic term $\lambda_3 Q_3\to 2\lambda_3 |\vec{\Delta}|^4$ does not lift the degeneracy of different real configurations. Instead, as laid out in Ref. \cite{BoettcherComplex}, the sextic terms in the free energy favour the uniaxial nematic state with $\vec{\Delta}=\Delta(0,1,0,0,0)$. For the weak coupling transition with $\lambda_1,\lambda_2>0$, the manifold of states $\vec{\Delta}$ satisfying $\vec{\Delta}^2=0$ is energetically degenerate for $\lambda_3=0$. If $\lambda_3>0$, $Q_3$ needs to be minimized which corresponds to $Q_3\to\frac{4}{3}|\vec{\Delta}|^4$ and $\vec{\Delta}=\frac{\Delta}{\sqrt{2}}(1,\rmi,0,0,0)$. In contrast, for $\lambda_3<0$, $Q_3$ needs to be maximized, which corresponds to $Q_3\to4|\vec{\Delta}|^4$ and $\vec{\Delta}=\frac{\Delta}{\sqrt{2}}(1,0,0,0,\rmi)$ \cite{PhysRevA.9.868}.

In the presence of cubic anisotropy, the Lagrangian in Eq. (\ref{qbt2}) receives an additional contribution
\begin{align}
 \label{qbt5} L_{\rm aniso} = \psi^\dagger \Bigl(\delta \cdot \sum_a  s_a d_a(-\rmi \nabla) \gamma_a \Bigr)\psi,
\end{align}
with dimensionless parameter $\delta\in [-1,1]$ and $s_{1,2}=-1$ and $s_{3,4,5}=+1$. (We follow the conventions of Ref. \cite{PhysRevB.95.075149}.) For nonzero $\delta$ the quadratic mass term for $\vec{\vphi}$ splits into two distinct contributions $r_{\rm E}$ and $r_{\rm T}$ for $\vec{\eta}$ and $\vec{\chi}$ as in Eq. (\ref{cub3}). Evaluating the expressions in Eqs. (\ref{ini15}) and (\ref{ini16}) we find that small $\delta<0$ favors $N=2$ fluctuating components, and small $\delta>0$ favors $N=3$ fluctuating components. These statements are true for both the strong-coupling and weak-coupling transitions, see Eqs. (\ref{ini18})-(\ref{ini25}). From the band structure of YPtBi given in Ref. \cite{2016arXiv160303375K} we determine $\delta=-0.19$ \cite{BoettcherComplex}, which is considerably small and negative, so that the number of fluctuating components is likely reduced to $N=2$. The critical theory is then described by the $\text{O}(2)\times \text{O}(2)$ symmetric effective action S$_{\rm E}$ given in Eq. (\ref{cub4}). The phase structure of this model has been discussed at the end of Sec. \ref{SecFlowN}. Note that due to $\lambda_{1,2}>0$, the couplings for the weak-coupling transition are within the domain of attraction of the proposed chiral fixed points for $N=2,3$ with $u_\star,v_\star>0$ \cite{PhysRevB.63.140414}, if they exist.

\section{Summary and Outlook}\label{SecDis}
In this work we have analyzed the critical properties of the complex tensor ordering phase transition in three dimensions. In the following we summarize our findings and point out directions along which the present calculations should be improved in future studies.

To capture the critical fluctuations of the complex tensor field we derived the low-energy effective action for the bosonic order parameter field and studied the scale dependence of its couplings by means of the perturbative RG. We observed that both the kinetic and interaction terms of the bosonic Lagrangian feature peculiar features that are not commonly encountered in $\text{O}(N)$-models for phase transitions. These features result both from the three-dimensionality of space and the irreducibility of the tensor field. Since they are intimately tied to the three-dimensional setting, the perturbative RG close to the critical dimension of four is somewhat problematic. Studying three complementary models for the critical field theory, however, we found qualitatively consistent results in all three of them, which gives us some confidence on their reliability (within the confines of perturbation theory).

For the isotropic system with $N=5$ fluctuating components we found that, although mean-field theory for a particular microscopic model of complex tensor order may predict a second-order phase transition, fluctuations of the order parameter induce a first-order phase transition. However, since the flow equations permit an unstable interacting fixed point, RG trajectories that are initiated in its vicinity might show a slowing down of the running of couplings, so that the corresponding transition is only weakly first-order and might appear continuous in experiment, with an extended temperature range that shows scaling behavior of observables. The absence of a weak-coupling fixed points in the plane spanned by $(\lambda_1,\lambda_2,\lambda_3)$ is in agreement with other studies on the $N$-component model $S'$, which do not find an attractive fixed point for $N=5$.

We have further shown that cubic anisotropy reduces the number of fluctuating components of the order parameter field to $N=2,3$ and that the associated critical field theories are identical to those of frustrated magnets. The critical properties of the latter, on the other hand, are still debated among the experts in the field, and the question on the order of the transition essentially remains open.

When the bosonic theory is derived from the underlying microscopic model for isotropic Luttinger semimetals, it describes critical phenomena at the putative second-order phase transitions towards complex tensor order derived in Ref. \cite{BoettcherComplex}. We have shown that both the strong-coupling and weak-coupling transitions of the model become of first order due to fluctuations, the second one, however, only weakly so. We have further shown that cubic anisotropy reduces the number of field components to $N=2,3$, and for YPtBi the microscopic parameters are such that $N=2$ and the flow is initialized for small $u_\star,v_\star>0$. It thus lies within the attractive domain of the proposed chiral fixed point of the $\text{O}(2)\times\text{O}(2)$ model and experiments on the second-order phase transition in the half-Heusler compound may thus contribute to the outstanding question of the critical physics of this model.

Our analysis indicates that the phase structure of complex tensor order is determined by the strong-coupling regime of the model, because both (i) the runaway flow of couplings and (ii) the potential existence of additional strong-coupling fixed points in the plane $(\lambda_1,\lambda_2,\lambda_3)$ is related to physics outside the perturbatively accessible domain. In order to improve on the results presented here it is thus necessary to apply genuinely nonperturbative methods such as classical Monte Carlo computations \cite{PelissettoVicari,PhysRevB.65.144520,PhysRevB.74.144506} or the Functional Renormalization Group (FRG) \cite{Wetterich1993,Berges:2000ew,Pawlowski20072831,Gies:2006wv,Delamotte:2007pf,Kopietz2010,Metzner:2011cw,Braun:2011pp,Boettcher:2012cm}. Both methods can explicitly resolve the first-order transition and the jump of the order parameter at the transition because they allow to track the RG flow of higher-order terms in the field beyond the quartic level and their applicability is, in principle, not limited to the weak-coupling regime. In particular, our derivation of the RG beta functions presented in App. \ref{AppRG} can be generalized to the local potential approximation of the exact FRG equation by means of a few suitable modification. A definite answer to the existence of the chiral fixed points in the cubic models for frustrated magnetism may eventually be obtained from the conformal bootstrap approach \cite{1126-6708-2008-12-031,PhysRevD.86.025022,KosPoland}.

In our application to the microscopic Luttinger semimetal we first integrated out the fermion fluctuations on the mean-field level and then incorporated bosonic fluctuations. This procedure is a good approximation for the corresponding weak-coupling transition, where the effect of bosonic fluctuations is small and we have a clear separation of energy scales where fermionic or bosonic contributions to the phase structure are important. At the strong-coupling transition, however, we find a strong renormalization of the quartic self-interaction with a divergence of the runaway flow for $b=b_0 \sim 1$. For this transition, it would thus be more reasonable to simultaneously integrate out fermionic and bosonic fluctuations to account for their mutual coupling and feedback. The FRG could be used as a tool to resolve this interplay as has been demonstrated in related models for relativistic \cite{1997MPLA...12.2287L,1997PhLB..393..387B,SCHAEFER2005479,PhysRevE.88.042141,BoettcherSarma,2016EPJC...76..472B,PhysRevD.94.025027} and non-relativistic field theories \cite{PhysRevLett.103.220602,PhysRevX.4.021012,PhysRevA.91.013610}, where the finite-temperature phase structure including both first- and second-order transitions was addressed. A generic obstacle of the simultaneous inclusion of fluctuations of fermions and bosons in this setting consists in the regulator dependence that sets the relative scale between the fermionic and the bosonic cutoff. This effect influences predictions for observables in any finite truncation, although the effect is typically small \cite{Schnoerr:2013bk,PhysRevA.89.053630,PAWLOWSKI2017165}.

\acknowledgements We thank Dietrich Roscher and Michael M. Scherer for inspiring discussions. IB acknowledges funding by the German Research Foundation (DFG) under Grant No. BO 4640/1-1. IB and IFH are supported by the NSERC of Canada.

\begin{appendix}

\section{Gell-Mann matrices}\label{AppGell}

In this section we give the explicit representation of real Gell-Mann matrices $\{\Lambda^a\}$ in three and four dimensions that is used in this work. Generally, the matrices are symmetric traceless and normalized such that $\mbox{tr}(\Lambda^a\Lambda^b)=2\delta_{ab}$. In $d$ dimensions, they are $d \times d$ matrices. They constitute an orthogonal basis for symmetric traceless matrices, and every symmetric traceless matrix  $M$ can be represented as $M=M_a\Lambda^a$ with $M_a=\frac{1}{2}\mbox{tr}(M\Lambda^a)$. In particular, this decomposition can also be applied if the matrix $M$ has complex entries.

In three dimensions, the five real Gell-Mann matrices read
\begin{align}
 \label{field3} \Lambda^1 &= \begin{pmatrix} 1 & 0 & 0 \\ 0 & -1 & 0 \\ 0 & 0 & 0 \end{pmatrix},\ \Lambda^2 = \frac{1}{\sqrt{3}} \begin{pmatrix} -1 & 0 & 0 \\ 0 & -1 & 0 \\ 0 & 0 & 2 \end{pmatrix},\\
 \nonumber \Lambda^3 &= \begin{pmatrix} 0 & 0 & 1 \\ 0 & 0 & 0 \\ 1 & 0 & 0 \end{pmatrix},\ \Lambda^4 = \begin{pmatrix} 0 & 0 & 0 \\ 0 & 0 & 1 \\ 0 & 1 & 0  \end{pmatrix},\ \Lambda^5 = \begin{pmatrix} 0 & 1 & 0 \\ 1 & 0 & 0 \\ 0 & 0 & 0 \end{pmatrix}.
\end{align}
Note that the first two matrices are diagonal, whereas the last three matrices are off-diagonal. In four dimensions, the corresponding nine matrices are 
\begin{align}
 \nonumber \Lambda^1 &= \begin{pmatrix} 1 & 0& 0& 0\\ 0& -1 &0 & 0\\ 0& 0& 0 &0 \\ 0& 0& 0& 0 \end{pmatrix},\ \Lambda^2 = \frac{1}{\sqrt{3}}\begin{pmatrix} -1 &0 &0 &0 \\0 & -1 & 0&0 \\ 0& 0& 2 & 0\\ 0& 0& 0& 0 \end{pmatrix},\\
 \nonumber \Lambda^3 &= \begin{pmatrix} 0 & 0 & 1 & 0\\ 0 & 0 & 0 & 0\\ 1 & 0 & 0 & 0\\ 0&0 &0 & 0\end{pmatrix},\ \Lambda^4 = \begin{pmatrix} 0 & 0 & 0 &0 \\ 0 & 0 & 1 &0 \\ 0 & 1 & 0 &0 \\0 & 0& 0& 0\end{pmatrix},\\
 \nonumber \Lambda^5 &= \begin{pmatrix} 0 & 1 & 0 &0 \\ 1 & 0 & 0 &0 \\ 0 & 0 & 0 & 0\\ 0& 0&0 & 0\end{pmatrix},\ \Lambda^6 = \begin{pmatrix}0 & 0&0& 1 \\ 0&0 & 0 &0  \\ 0& 0 & 0&0 \\ 1 &0 &0 & 0\end{pmatrix},\\ 
 \nonumber \Lambda^7 & = \begin{pmatrix} 0 &0 &0 &0 \\0 & 0 &0 &0 \\ 0&0 & 0 & 1 \\ 0& 0& 1 & 0 \end{pmatrix},\ \Lambda^8 =\begin{pmatrix} 0 &0 &0 &0 \\ 0& 0 & 0 & 1 \\ 0& 0 & 0 & 0 \\0 & 1 & 0 & 0\end{pmatrix},\\
 \label{gell1} \Lambda^9 &= \frac{1}{\sqrt{6}} \begin{pmatrix} -1 &0 &0& 0\\ 0& -1 &0 &0 \\ 0& 0& -1 &0\\ 0& 0&0 & 3 \end{pmatrix}.
\end{align}

We define the $J$-symbols through Eq. (\ref{field4}). Note that since the trace is cyclic, $J_{ab\dots c}$ is invariant under a cyclic permutation of its indices. Furthermore, since $\mbox{tr}(A)=\mbox{tr}(A^{\rm T})$ and the matrices $\{\Lambda^a\}$ being symmetric we have
\begin{align}
 \nonumber J_{ab\dots cd} &= \mbox{tr}(\Lambda^a \Lambda^b \cdots \Lambda^c \Lambda^d)= \mbox{tr}([\Lambda^a \Lambda^b \cdots \Lambda^c \Lambda^d] ^{\rm T})\\
 \label{gell2} &= \mbox{tr}(\Lambda^d \Lambda^c \cdots \Lambda^b \Lambda^a) = J_{dc\dots ba},
\end{align}
so that the $J$'s are also invariant under reversing the order of their indices. These two operations exhaust all possible permutations of three indices and therefore the order of indices in $J_{abc}$ is irrelevant.

Most values of $J_{abc}$ vanish. In three dimensions the nonvanishing ones are
\begin{align}
 \nonumber J_{112}&=-\frac{2}{\sqrt{3}},\ J_{133}=1,\ J_{144}=-1,\ J_{222}=\frac{2}{\sqrt{3}},\ \\
 \label{gell3} J_{233}&=\frac{1}{\sqrt{3}},\ J_{244}=\frac{1}{\sqrt{3}},\ J_{255}=-\frac{2}{\sqrt{3}},\ J_{345}=1.
\end{align}
Note that $J_{abc} \propto \int_{\textbf{q}} d_a(\textbf{q})d_b(\textbf{q})d_c(\textbf{q})$. See App. C of Ref. \cite{PhysRevB.93.205138} for a general discussion of properties of Gell-Mann matrices and $d_a$-functions in $d$ dimensions.

\section{Relation to $\text{U}(3)\times \text{U}(3)$ symmetric matrix models}\label{AppMat}
In this appendix we clarify the difference of our model (and the corresponding RG flow) to the $\text{U}(3)\times\text{U}(3)$ symmetric matrix model of Refs. \cite{PhysRevD.29.338,PhysRevB.23.3549,BERGES1997675} more explicitly. For this consider an arbitrary $3\times 3$ complex matrix $\Phi$ described by the effective action
\begin{align}
 \nonumber \tilde{S}[\Phi] = \int\mbox{d}^3x \Bigl[{}& \frac{1}{2}\mbox{tr}(\nabla \Phi^\dagger\cdot \nabla\Phi) + \frac{u_1}{4}[\mbox{tr}(\Phi^\dagger\Phi)]^2 \\
 \label{mat1} &+ u_3 \mbox{tr}(\Phi^\dagger\Phi\Phi^\dagger\Phi)\Bigr].
\end{align}
Then $\tilde{S}$ is invariant under the  $\text{U}(3)\times\text{U}(3)$ transformation $\Phi \to \Phi'=U\Phi V^\dagger$ with $U,V\in\text{U}(3)$. Furthermore, since a term $u_2 |\mbox{tr}(\Phi^2)|^2$ breaks this symmetry, the coupling $u_2$ cannot be generated from $u_{1,3}$ if it is initially absent: $u_2=0 \Longrightarrow \dot{u}_2=0$.

At first sight, $\tilde{S}$ appears to correspond to the theory described by $S$ in Eq. (\ref{flow0}) when $\lambda_2=0$. This, however, is not the case since for a symmetric traceless matrix $\phi$, the transformed matrix $\phi'=U \phi V^\dagger$ is no longer symmetric and traceless, so that the $\text{U}(3)\times\text{U}(3)$ transformation was not an admissible operation. To make the situation more transparent we parametrize $\Phi$ by its $N=9$ complex components according to
\begin{align}
\label{mat3b} \Phi=\sum_{a=1}^9 \zeta_a\Sigma^a,
\end{align}
where $\Sigma^a=\Lambda^a$ for $a=1,\dots,5$ with the real $3\times 3$ Gell-Mann matrices from Eq. (\ref{field3}), $\Sigma^6=\sqrt{\frac{2}{3}}\mathbb{1}_3$, and we employ the additional imaginary Gell-Mann matrices
\begin{align}
 \label{mat4} \Sigma^7 = \begin{pmatrix} 0 & 0 & \rmi \\ 0 & 0 & 0 \\ -\rmi & 0 & 0 \end{pmatrix},\ \Sigma^8 =\begin{pmatrix} 0 & 0 & 0 \\ 0 & 0 & \rmi \\ 0 & -\rmi & 0 \end{pmatrix},\ \Sigma^9 &= \begin{pmatrix} 0 & \rmi &0 \\ -\rmi & 0 & 0 \\ 0 & 0 & 0 \end{pmatrix}.
\end{align}
The completeness of the set $\{\Sigma^a\}$ to represent any complex $3\times 3$ matrix results from the fact that we can uniquely decompose each such matrix into a diagonal, symmetric-traceless, and antisymmetric-traceless part. Introduce further the $H$-symbols
\begin{align}
 \label{mat5} H_{ab\dots c} = \mbox{tr}(\Sigma^a\Sigma^b\cdots\Sigma^c).
\end{align}
The action and partition function for the $\text{U}(3)\times\text{U}(3)$ symmetric model are then given by
\begin{align}
 \nonumber \tilde{S}[\zeta] = \int\mbox{d}^3x\Bigl({}&\zeta_a^*(-\nabla^2)\zeta_a + u_1(\zeta_a^*\zeta_a)^2\\
 \label{mat6} &+u_3 H_{abcd}\zeta_a^*\zeta_b\zeta_c^*\zeta_d\Bigr)
\end{align}
and
\begin{align}
 \label{mat7} \tilde{Z}= \int\mbox{D}\zeta \mbox{D}\zeta^* e^{-\tilde{S}[\zeta]},
\end{align}
respectively. In contrast, $S$ purely in terms of $\vphi_a$ reads
\begin{align}
 \nonumber S[\vphi] =\int\mbox{d}^3x\Bigl({}&\vphi_a^*(-\nabla^2)\vphi_a + \lambda_1(\vphi_a^*\vphi_a)^2+\lambda_2|\vphi_a\vphi_a|^2\\
 \label{mat2} &+\lambda_3 J_{abcd}\vphi_a^*\vphi_b\vphi_c^*\vphi_d\Bigr)
\end{align}
with partition function
\begin{align}
 \label{mat3} Z = \int\mbox{D}\vphi \mbox{D}\vphi^* e^{-S[\vphi]}.
\end{align}
The irreducibility of $\phi$ is fully incorporated by the $J_{abcd}$-vertex in the quartic term of $S$. In this representation it is then obvious that the $\text{SO}(3)\times \text{U}(1)$ and $\text{U}(3)\times \text{U}(3)$ symmetric models differ in two crucial aspects: First, the former comprises five fluctuating complex components, the latter comprises nine fluctuating complex components. Second, they explicitly differ in the vertices $\lambda_3 J_{abcd}$ and $u_3 H_{abcd}$ that enter the quartic terms $\lambda_3 \mbox{tr}(\phi^\dagger\phi\phi^\dagger\phi)$ and $u_3 \mbox{tr}(\Phi^\dagger\Phi\Phi^\dagger\Phi)$, respectively.

As a proof of principle we derive the flow equations for the $\text{U}(N_{\rm f})\times \text{U}(N_{\rm f})$ symmetric model for $N_{\rm f}=3$ by means of the procedure outlined in App. \ref{AppRG}. For this we replace $\bar{\lambda}_1\to u_1$, $\bar{\lambda}_2\to u_2=0$, $\bar{\lambda}_3\to u_3$, and $J_{abcd} \to H_{abcd}$ in Eq. (\ref{rg15}). Note, however, that the identities for the $J$-symbols given below Eq. (\ref{rg15}) are no longer valid, so that the full expressions have to be inserted, that is $2J_{acbd}\to H_{acbd}+H_{cbda}$ and $4J_{abcd} \to H_{abcd}+H_{adcb}+H_{cbad}+H_{cdab}$. Eqs. (\ref{rg12})-(\ref{rg14}) remain valid. Summing over nine complex components and rescaling the couplings by a constant prefactor we obtain
\begin{align}
 \label{mat8} \dot{u}_1 &=\vare u_1 -\frac{13}{3}u_1^2-4u_1u_3-u_3^2,\\
 \label{mat9} \dot{u}_2 &=0,\\
 \label{mat10} \dot{u}_3 &= \vare u_3 -2u_1u_3-2u_3^2.
\end{align}
This agrees for $N_{\rm f}=3$ with the expressions
\begin{align}
  \label{mat11} \dot{u}_1 &=\vare u_1 -\frac{N_{\rm f}^2+4}{3}u_1^2-\frac{4N_{\rm f}}{3}u_1u_3-u_3^2,\\
 \label{mat12} \dot{u}_3 &= \vare u_3 -2u_1u_3-\frac{2N_{\rm f}}{3}u_3^2.
\end{align}
given in Ref. \cite{PhysRevD.29.338}.

\section{Inequalities for tensor invariants}\label{AppIneq}
In this section we derive the inequalities (\ref{field22}). The first two ones, $0\leq |\vec{\vphi}^2|^2\leq |\vec{\vphi}|^4$, are obviously true. For the remaining two ones note first that every square matrix $\phi$ can be written in polar decomposition as
\begin{align}
 \label{ineq1} \phi = HU,\ H\ \text{Hermitean},\ U\ \text{unitary}.
\end{align}
We briefly recall the proof of this fact. First note that $\phi\phi^\dagger$ is Hermitean and non-negative, so there is a diagonal matrix $D^2$ with non-negative entries, and a unitary $V$, such that $\phi \phi^\dagger=V D^2 V^\dagger =(V D V^\dagger)(V D V^\dagger)=H^2$. The matrix $H:=V DV^\dagger$ is Hermitean and positive semi-definite.Typically $\phi$ (and thus $H$) will be invertible. In these cases define $U:=H^{-1}\phi $, which is unitary.  In general, even for singular $\phi$, the singular value decomposition guarantees that there are unitary $V,W$ such that $\phi=VDW^\dagger$, and consequently $U=VW^\dagger $.

The Hermitean matrix $H$ is characterized by three non-negative eigenvalues $(h_1,h_2,h_2)$. We have
\begin{align}
 \label{ineq2} |\vec{\vphi}|^4 = \frac{1}{4}[\mbox{tr}(\phi^\dagger\phi)]^2 = \frac{1}{4}[\mbox{tr}(H^2)]^2 = \frac{1}{4}[h_1^2+h_2^2+h_3^2]^2.
\end{align}
For the remainder of this section we normalize $|\vec{\vphi}|^4=1$ so that $h_1^2+h_2^2+h_3^2=2$. We then find
\begin{align}
 \nonumber \mbox{tr}(\phi^\dagger\phi\phi^\dagger\phi) &= \mbox{tr}(H^4) = (h_1^4+h_2^4+h_3^4) \\
 \label{ineq3} &= (h_1^4+h_2^4+[2-h_1^2-h_2^2]^2).
\end{align}
The minima and maxima of this expression can readily be determined in the planar area $0\leq h_1^2+h_2^2\leq 2$. For $(h_1,h_2,h_3)=\sqrt{2/3}(1,1,1)$ we have minimal value $\mbox{tr}(\phi^\dagger\phi\phi^\dagger\phi)=\frac{4}{3}$. The maximal value $\mbox{tr}(\phi^\dagger\phi\phi^\dagger\phi)=4$ is obtained when two of the $h_i$ vanish, say $(h_1,h_2,h_3)=(0,0,\sqrt{2})$. In the derivation of the minimal and maximal values for this quartic invariant we only used the decomposition in Eq. (\ref{ineq1}), so that the finding is true for any square $3\times 3$ matrix not necessarily symmetric or traceless.

Note that in the same manner the maximal and minimal values of the sextic invariant $|\mbox{tr}(\phi^3)|^2$ can be derived. Since for every symmetric traceless $3\times 3$ matrix we have
\begin{align}
 \label{ineq4} \mbox{tr}(\phi^3) = 3 \mbox{det}(\phi),
\end{align}
we have
\begin{align}
 \label{ineq5} |\mbox{tr}(\phi^3)|^2 = 9 h_1^2h_2^2h_3^2=9 h_1^2h_2^2(2-h_1^2-h_2^2).
\end{align}
We used here $\mbox{det}(HU)=\mbox{det}(H)\mbox{det}(U)$ and $|\mbox{det}(U)|=1$ for $U$ unitary. The maximal value $|\mbox{tr}(\phi^3)|^2=\frac{8}{3}$ is obtained for $(h_1,h_2,h_3)=\sqrt{2/3}(1,1,1)$. The obvious minimal value $|\mbox{tr}(\phi^3)|^2=0$ is obtained for $\phi$ being singular (i.e. non-invertible), which means that at least one of the $h_i$ vanishes.

\section{Derivation of renormalization group equations}\label{AppRG}

In order to compute the running of the quartic couplings we employ here the general formula for the one-loop correction to the effective potential, $\Delta U(\phi)$, which constitutes an economic way to access the beta functions for $K\neq 0$. The running of couplings can equivalently be derived from standard perturbation theory \cite{herbutbook}. We also performed this approach with agreeing results, but will not present it here. We define the effective potential as the part of the effective Lagrangian that does not depend on derivatives of the field. A saddle-point expansion of the path integral (see for instance the lecture notes \cite{Boettcher:2012cm}) then yields the ``trace-log formula''
\begin{align}
 \label{rg1} \Delta U(\phi) = \frac{1}{2} \mbox{tr} \int_{\textbf{q}}^\prime \ln \mathcal{G}^{-1}(\phi,\textbf{q}),
\end{align}
where $\mathcal{G}^{-1}(\phi,\textbf{q})$ is the inverse perturbative propagator in the presence of a spatially constant background field $\phi$. It is related to the second functional derivative of $S[\phi]$ by means of
\begin{align}
 \nonumber  \frac{\delta^2 S}{\delta \vphi_a(\textbf{q}) \delta \vphi_b(\textbf{q}')}[\phi(\textbf{x})=\phi] =\mathcal{G}_{\vphi\vphi,ab}^{-1}(\phi,\textbf{q}) \delta^{(d)}(\textbf{q}-\textbf{q}'),\\
 \label{rg2} \frac{\delta^2 S}{\delta \vphi_a^*(\textbf{q}) \delta \vphi_b(\textbf{q}')}[\phi(\textbf{x})=\phi] =\mathcal{G}_{\vphi^*\vphi,ab}^{-1}(\phi,\textbf{q}) \delta^{(d)}(\textbf{q}-\textbf{q}')
\end{align}
and
\begin{align}
 \mathcal{G}^{-1} = \begin{pmatrix} \mathcal{G}^{-1}_{\vphi\vphi} & \mathcal{G}^{-1}_{\vphi\vphi^*} \\ \mathcal{G}^{-1}_{\vphi^*\vphi} & \mathcal{G}^{-1}_{\vphi^*\vphi^*}\end{pmatrix}.
\end{align}
The trace in Eq. (\ref{rg1}) sums over all internal field degrees of freedom, which here comprises $(\vphi_a,\vphi_a^*)$. The presence of the background field allows us to conveniently take derivatives of $\Delta U(\phi)$, which, in this way, serves as generating function for the one-loop corrections of all couplings of the effective potential. For instance, we have
\begin{align}
 \label{rg3} \frac{\partial \Delta U}{\partial \vphi_a}(\phi) = \frac{1}{2}\mbox{tr} \int_{\textbf{q}}^\prime \mathcal{G}(\phi,\textbf{q}) \frac{\partial \mathcal{G}^{-1}(\phi,\textbf{q})}{\partial \vphi_a}.
\end{align}
We define the momentum-shell integration by means of
\begin{align}
 \label{rg4} \int_{\textbf{q}}^\prime (\dots) = \frac{1}{(2\pi)^d} \int_{\Omega/b}^\Omega \mbox{d}q q^{d-1} \int_{\mathbb{S}^{d-1}}(\dots).
\end{align}
The angular momentum integration will be carried out in either three or four dimensions, depending on the model.

In order to compute the running of the quartic couplings we first determine
\begin{align}
 \label{rg6} I_1 &= \frac{\partial^4 \Delta U}{\partial \vphi_1^*\partial \vphi_1^*\partial \vphi_1\partial \vphi_1}\Bigr|_{\phi=0},\\
 \label{rg7} I_2 &= \frac{\partial^4 \Delta U}{\partial \vphi_1^*\partial \vphi_1\partial \vphi_2^*\partial \vphi_2}\Bigr|_{\phi=0},\\
 \label{rg8} I_3 &= \frac{\partial^4 \Delta U}{\partial \vphi_1^*\partial \vphi_3^*\partial \vphi_4\partial \vphi_5}\Bigr|_{\phi=0}.
\end{align}
Writing $U = \bar{\lambda}_1 |\vec{\vphi}|^4 + \bar{\lambda}_2 |\vec{\vphi}^2|^2 + \bar{\lambda}_3 \mbox{tr}(\phi^\dagger\phi\phi^\dagger\phi)$ we identify
\begin{align}
 \label{rg9} I_1 &= 4\Delta \bar{\lambda}_1 + 4\Delta \bar{\lambda}_2 + 8\Delta \bar{\lambda}_3,\\
 \label{rg10} I_2 &= 2\Delta\bar{\lambda}_1 +\frac{8}{3}\Delta\bar{\lambda}_3,\\
 \label{rg11} I_3 &= -4\Delta \bar{\lambda}_3,
\end{align}
where $\Delta \bar{\lambda}_m$ is the one-loop correction to $\bar{\lambda}_m$. The linear set of equations (\ref{rg9})-(\ref{rg11}) can be inverted and we eventually obtain
\begin{align}
 \label{rg12} \Delta \bar{\lambda}_1 &= \frac{1}{2}I_2 +\frac{1}{3}I_3,\\
 \label{rg13} \Delta \bar{\lambda}_2 &= \frac{1}{4}I_1 -\frac{1}{2} I_2 +\frac{1}{6} I_3,\\
 \label{rg14} \Delta \bar{\lambda}_3 &= -\frac{1}{4}I_3.
\end{align}
Since the one-loop correction depends on $b$, we can define running couplings $\bar{\lambda}_m(b) = \bar{\lambda}_m(1) + \Delta \bar{\lambda}_m(b)$ and the renormalization group flow $\dot{\bar{\lambda}}_m = \mbox{d} \bar{\lambda}_m(b) / \mbox{d}\ln b$. As we increase $b$, we successively lower the cutoff of the theory and thereby include fluctuations at high momenta to obtain an effective infrared theory.

We first derive the flow equations for the quartic couplings for $K=0$. We have 
\begin{align}
 \nonumber \mathcal{G}_{ab}^{-1}(\phi,\textbf{q}) &= \begin{pmatrix} 2\bar{\lambda}_2 (\vec{\vphi}^2)^*\delta_{ab} & (q^2+2\bar{\lambda}_1|\vec{\vphi}|^2)\delta_{ab} \\ (q^2+2\bar{\lambda}_1|\vec{\vphi}|^2)\delta_{ab} & 2\bar{\lambda}_2(\vec{\vphi}^2)\delta_{ab} \end{pmatrix}\\
 \nonumber & +\begin{pmatrix} 2\bar{\lambda}_1 \vphi_a^*\vphi_b^* & 2\bar{\lambda}_1\vphi_a^*\vphi_b + 4\bar{\lambda}_2 \vphi_a\vphi_b^* \\ 2\bar{\lambda}_1 \vphi_a\vphi_b^*+4\bar{\lambda}_2 \vphi_a^*\vphi_b & 2\bar{\lambda}_1 \vphi_a\vphi_b \end{pmatrix}\\
 \label{rg15} &+\begin{pmatrix} 2\bar{\lambda}_3\vphi_c^*\vphi_d^*J_{acbd} & 4\bar{\lambda}_3 \vphi_c\vphi_d^* J_{abcd} \\ 4\bar{\lambda}_3 \vphi_c^*\vphi_dJ_{abcd} & 2\bar{\lambda}_3 \vphi_c\vphi_d J_{acbd} \end{pmatrix}.
\end{align}
In the last line we employed $J_{cadb}+J_{cbda}=2J_{acbd}$ and $J_{abcd}+J_{adcb}+J_{cbad}+J_{cdab}=4J_{abcd}$ which follows from the fact that we can cyclic permute the indices of $J_{ab\dots c}$ and reverse their order, see App. \ref{AppGell}. We have
\begin{align}
 \nonumber \frac{\partial^4 \Delta U}{\partial \vphi_a^*\partial \vphi_b^*\partial \vphi_c\partial \vphi_d}\Bigr|_{\phi=0} = &{}-\frac{1}{2}\mbox{tr}\int_{\textbf{q}}^\prime G \frac{\partial^2 \mathcal{G}^{-1}}{\partial \vphi_a^*\partial \vphi_b^*} G \frac{\partial^2 \mathcal{G}^{-1}}{\partial \vphi_c\partial \vphi_d}\\
 \nonumber & -\frac{1}{2}\mbox{tr}\int_{\textbf{q}}^\prime G \frac{\partial^2 \mathcal{G}^{-1}}{\partial \vphi_a^*\partial \vphi_c} G \frac{\partial^2 \mathcal{G}^{-1}}{\partial \vphi_b^*\partial\vphi_d} \\
 \label{rg16} &-\frac{1}{2}\mbox{tr}\int_{\textbf{q}}^\prime G \frac{\partial^2 \mathcal{G}^{-1}}{\partial \vphi_a^*\partial \vphi_d}G \frac{\partial^3 \mathcal{G}^{-1}}{\partial \vphi_b^*\partial \phi_c}
\end{align}
with $G=G(\textbf{q})=\mathcal{G}(\phi=0,\textbf{q})$. We used that the field dependence of $\mathcal{G}^{-1}$ is purely quadratic in the components $\vphi_a$, so that derivatives of $\mathcal{G}^{-1}$ with respect to three or more fields vanish identically.  For $\phi=0$ we can easily invert $\mathcal{G}^{-1}$ to obtain
\begin{align}
 \label{rg19} G_{ab}(\textbf{q}) = \frac{1}{q^2}\begin{pmatrix} 0 & \delta_{ab} \\ \delta_{ab} & 0 \end{pmatrix}.
\end{align}
Furthermore, given the explicit expression in Eq. (\ref{rg15}), it is easy to determine the second derivatives of $\mathcal{G}^{-1}$ and eventually perform the contractions by hand or with Mathematica. We arrive at
\begin{align}
 \nonumber I_1 &= -8\Bigl(9\bar{\lambda}_1^2+10\bar{\lambda}_1\bar{\lambda}_2+9\bar{\lambda}_2^2+\frac{116}{3}\bar{\lambda}_1\bar{\lambda}_3+\frac{52}{3}\bar{\lambda}_2\bar{\lambda}_3\\
 \label{rg20} &+\frac{452}{9}\bar{\lambda}_3^2\Bigr)\int_{\textbf{q}}^\prime \frac{1}{q^4},\\
 \nonumber I_2 &= -4\Bigl(9\bar{\lambda}_1^2+4\bar{\lambda}_1\bar{\lambda}_2+4\bar{\lambda}_2^2+\frac{104}{3}\bar{\lambda}_1\bar{\lambda}_3 +\frac{32}{3}\bar{\lambda}_2\bar{\lambda}_3\\
 \label{rg21}&+\frac{128}{3}\bar{\lambda}_3^2\Bigr)\int_{\textbf{q}}^\prime \frac{1}{q^4},\\
 \label{rg22} I_3 &= \frac{16}{3}\bar{\lambda}_3(9\bar{\lambda}_1-6\bar{\lambda}_2+31\bar{\lambda}_3)\int_{\textbf{q}}^\prime \frac{1}{q^4}.
\end{align}
Using $\int_{\textbf{q}}^\prime \frac{1}{q^4} = \frac{1}{2\pi^2}\Omega^{-\vare}\ln b$ for $\vare \to 0$, rescaling the couplings according to $\lambda_m=\frac{1}{2\pi^2}\Omega^{-\vare} \bar{\lambda}_m$,  we arrive at the one-loop flow equations (\ref{flow1})-(\ref{flow3}).

Now we determine the flow equations for arbitrary $K$. The sole change in the formulas is that we have to replace $q^2\delta_{ab}$ in the off-diagonal of $\mathcal{G}^{-1}$ in Eq. (\ref{rg15}) by
\begin{align}
 \label{rg23} D_{ab}(\textbf{q}) = q^2 \delta_{ab}+\frac{K}{\sqrt{3}} J_{abc} d_c(\textbf{q})
\end{align}
from Eq. (\ref{field20}). Since this term is independent of the fields, it does not effect the vertices $\partial^2\mathcal{G}^{-1}/\partial \vphi^2$ that enter the expressions for $I_{1,2,3}$. However, the additional kinetic term contributes to the propagator. We have
\begin{align}
 \label{rg24} G(\textbf{q}) = \begin{pmatrix} 0 & D(\textbf{q})^{-1} \\ D(\textbf{q})^{-1} & 0 \end{pmatrix},
\end{align}
where $D(\textbf{q})^{-1}$ is the inverse of the $5\times 5$ matrix $D(\textbf{q})$. Here we can employ the general formula for the inverse of an invertible $5\times 5$ matrix $D$ given by 
\begin{align}
 \nonumber D^{-1} = &{}\frac{1}{\mbox{det}(D)} \Biggl(\frac{1}{24}\Bigl[(\mbox{tr} D)^4-6(\mbox{tr}D)^2\mbox{tr}(D^2)+3[\mbox{tr}(D^2)]^2\\
 \nonumber & +8(\mbox{tr}D)\mbox{tr}(D^3)-6\mbox{tr}(D^4)\Bigr]\mathbb{1}_5\\
 \nonumber &-\frac{1}{6}[(\mbox{tr} D)^3-3(\mbox{tr} D)\mbox{tr}(D^2)+2\mbox{tr}(D^3)]D\\
 \label{rg25} &+\frac{1}{2}[(\mbox{tr}D)^2-\mbox{tr}(D^2)]D^2-(\mbox{tr}D)D^3+D^4\Biggr).
\end{align}
(The formula follows from the Cayley--Hamilton theorem.) For $D=D(\textbf{q})$ from Eq. (\ref{rg23}) we obtain
\begin{align}
 \label{rg26a} \mbox{det}(D) &= \frac{1}{243}(3+2K)(3-2K)^2(3+K)^2q^{10},\\
 \label{rg26} \mbox{tr}(D) &= 5q^2,\\
 \label{rg27} \mbox{tr}(D^2) &= \frac{1}{9}(45+14K^2)q^4,\\
 \label{rg28} \mbox{tr}(D^3) &= \frac{1}{9}(45+42K^2-2K^3)q^6,\\
 \label{rg29} \mbox{tr}(D^4) &= \frac{1}{81} (405+756K^2 -72K^3+50 K^4)q^8.
\end{align}
The angular dependence introduced by the term proportional to $K$ makes the loop integration more involved. However, since the determinant is rotationally invariant, no angular dependence appears in the denominators. Consequently, the angular integration can be carried out explicitly and yields simple rational coefficients in the beta functions. The full beta functions for $K\neq 0$ are found to be
\begin{widetext}
 \begin{align}
 \nonumber \dot{\lambda}_1 =&{} \vare \lambda_1 + \frac{1}{35(3+2K)^2(3-2K)^2(3+K)^2}\Bigl[-18(25515+17010K+1953K^2+3372K^3+2228K^4)\lambda_1^2\\
 \nonumber &-72(2835+1890K+1197K^2+1308K^3+484K^4)\lambda_1\lambda_2-216(945+630K-189K^2-124K^3+12K^4)\lambda_2^2\\
 \nonumber &-288(4725+3150K+966K^2+1200K^3+592K^4)\lambda_1\lambda_3-96(8505+5670K+1197K^2+1644K^3+772K^4)\lambda_2\lambda_3\\
 \label{rg30} &-16(48195+32130K+19089K^2+21036K^3+10004K^4)\lambda_3^2\Bigr],\\
 \nonumber \dot{\lambda}_2 =&{}\vare \lambda_2 + \frac{1}{35(3+2K)^2(3-2K)^2(3+K)^2}\Bigl[-36K^2(189+180K+52K^2)\lambda_1^2\\
 \nonumber &-108(2835+1890K-231K^2-52K^3+116K^4)\lambda_1\lambda_2-18(14175+9450K+2961K^2+3660K^3+1748K^4)\lambda_2^2\\
 \nonumber &+48K^2(441+420K+52K^2)\lambda_1\lambda_3-24(19845+13230K-1449K^2-204K^3+716K^4)\lambda_2\lambda_3\\
 \label{rg31} &+16(19845+13230K+3024K^2+4056K^3+1216K^4)\lambda_3^2\Bigr],\\
 \nonumber \dot{\lambda}_3 =&{}\vare \lambda_3 + \frac{1}{35(3+2K)^2(3-2K)^2(3+K)^2}\Bigl[-162K^2(3+2K)(7+2K)(\lambda_1^2-4\lambda_1\lambda_2+2\lambda_2^2)\\
 \nonumber &-108(2835+1890K-189K^2-12K^3+92K^4)\lambda_1\lambda_3+216(3+K)(315+105K+52K^3)\lambda_2\lambda_3\\
 \label{rg32} &-36(29295+19530K-1029K^2+756K^3+1052K^4)\lambda_3^2\Bigr].
\end{align}
When expanding to quadratic order in $K$, we obtain the Eqs. (\ref{flow4})-(\ref{flow6}). Note that since the beta functions diverge as $|K|\to 3/2$ the fixed point couplings vanish accordingly to yield a solution of $\dot{\lambda}_i=0$. This behavior is visible in Fig. \ref{FigFixA}.

The computation of the flow equations for the four-dimensional model $S_{d=4}$ in Eq. (\ref{eps1}) follows the same steps as for the three-dimensional model. We therefore only discuss the necessary modifications. First note that in order to project onto the four couplings $\bar{g}_m$ in 
\begin{align}
 \label{rg33} U = \bar{g}_1 |\vec{\vphi}|^4 + \bar{g}_2 |\vec{\vphi}^2|^2 +\bar{g}_3 \mbox{tr}(\phi^\dagger\phi\phi^\dagger\phi) +\bar{g}_4\mbox{tr}(\phi^\dagger{}^2\phi^2),
\end{align}
we need four terms $I_m$. We choose
\begin{align}
 \label{rg34} I_1 &= \frac{\partial^4\Delta U}{\partial \vphi^*_1\partial\vphi_1^*\partial \vphi_1\partial \vphi_1}\Bigr|_{\phi=0} = 4\Delta \bar{g}_1+4\Delta \bar{g}_2+8\Delta \bar{g}_3+8\Delta \bar{g}_4,\\
 \label{rg35} I_2 &= \frac{\partial^4\Delta U}{\partial \vphi_1^*\partial\vphi_1\partial\vphi_2^*\partial \vphi_2}\Bigr|_{\phi=0} = 2\Delta \bar{g}_1+\frac{8}{3}\Delta \bar{g}_3+\frac{8}{3}\Delta \bar{g}_4,\\
 \label{rg36} I_3 &=\frac{\partial^4\Delta U}{\partial \vphi_1^*\partial \vphi_3^*\partial\vphi_4\partial \vphi_5}\Bigr|_{\phi=0} = -4\Delta \bar{g}_3+2\Delta \bar{g}_4,\\
 \label{rg37}  I_4 &= \frac{\partial^4\Delta U}{\partial \vphi_3^*\partial \vphi_7^*\partial \vphi_1\partial \vphi_6}\Bigr|_{\phi=0} = 2\Delta \bar{g}_4,
\end{align}
which leads us to
\begin{align}
 \label{rg38} \Delta \bar{g}_1 &= \frac{1}{2}I_2+\frac{1}{3}I_3-I_4,\\
 \label{rg39} \Delta \bar{g}_2 &= \frac{1}{4}I_1-\frac{1}{2}I_2+\frac{1}{6}I_3-\frac{1}{2}I_4,\\
 \label{rg40} \Delta \bar{g}_3 &= -\frac{1}{4}I_3 +\frac{1}{4}I_4,\\
 \label{rg41} \Delta \bar{g}_4 &= \frac{1}{2}I_4.
\end{align}
In Eq. (\ref{rg15}) we add a term proportional to $g_4$ according to
\begin{align}
 \mathcal{G}^{-1}_{ab}(\phi,\textbf{q}) = \begin{pmatrix} S^{(2)}_{\vphi_a\vphi_b} & [S^{(2)}_{\vphi^*_a\vphi_b}]^* \\ S^{(2)}_{\vphi^*_a\vphi_b} & [S^{(2)}_{\vphi_a\vphi_b} ]^*\end{pmatrix},
\end{align}
with
\begin{align}
 S^{(2)}_{\vphi^*_a\vphi_b} &= (q^2+2\bar{g}_1|\vec{\vphi}|^2)\delta_{ab}+2\bar{g}_1\vphi_a\vphi_b^*+4\bar{g}_2\vphi_a^*\vphi_b+4\bar{g}_3\vphi_c^*\vphi_dJ_{abcd}+2\bar{g}_4\vphi_c^*\vphi_d(J_{acbd}+J_{acdb}),\\
 S^{(2)}_{\vphi_a\vphi_b} &= 2\bar{g}_1\vphi_a^*\vphi_b^*+2\bar{g}_2\delta_{ab}(\vec{\vphi}^2)^*+2\bar{g}_3\vphi_c^*\vphi_d^*J_{acbd}+\bar{g}_4\vphi_c^*\vphi_d^*(J_{abcd}+J_{bacd}).
\end{align}
The computation of $G(\textbf{q})$ requires to invert the $9\times 9$ matrix
\begin{align}
\label{rg42} D_{ab}(\textbf{q}) = q^2\delta_{ab} + K\sqrt{\frac{3}{8}}J_{abc}d_c(\vec{q}) =: q^2\delta_{ab}+\delta D_{ab}(\textbf{q}).
\end{align}
Analogous to Eq. (\ref{rg25}), the general formula for the inverse of a $9\times 9$ matrix $D$ is of the form $D^{-1}=\frac{1}{\mbox{det}(D)}\sum_{k=0}^8 a_k D^k$, where the coefficients $k$ are functions of $\mbox{tr}(D^n)$ with $n\leq 8$. 
We expand $D^{-1}$ to quadratic order in $K$. For this we employ
\begin{align}
 \label{rg44} D^ n = (q^2\mathbb{1}_9+\delta D)^n = q^{2n}\mathbb{1}_9  + n q^{2(n-1)}\delta D + \frac{n(n-1)}{2}q^{2(n-2)} \delta D^2 + \mathcal{O}(K^3).
\end{align}
Now since $\mbox{tr}(\delta D)=0$ this implies
\begin{align}
 \nonumber \mbox{tr}(D^n) &= 9q^{2n} +\frac{n(n-1)}{2}q^{2(n-2)} \mbox{tr}(\delta D^2) +\mathcal{O}(K^3)\\
 \label{rg45}&= 3\Bigl(3+\frac{1}{2}n(n-1)K^2\Bigr)q^{2n}+\mathcal{O}(K^3),
\end{align}
where we used $\mbox{tr}(\delta D^2)=3K^2q^4$. Further
\begin{align}
 \label{rg46}\mbox{det}(D) = \Bigl(1-\frac{3}{2}K^2\Bigr)q^{18} +\mathcal{O}(K^3).
\end{align}
We observe that the absence of terms linear in $K$ in $D^{-1}$ is related to $\mbox{tr}(\delta D)=0$, which results from $\sum_a J_{aab}=0$. Since the latter equation holds for any dimension $d$ \cite{PhysRevB.93.205138}, the leading contributions to the beta functions in any $d$ and to any perturbative loop order is $\mathcal{O}(K^2)$. (Note that the determinant can also be expression in terms of $\mbox{tr}(D^n)$ due to the Cayley--Hamilton theorem.) Performing the angular integrals involved in $I_m$ in Eqs. (\ref{rg34})-(\ref{rg37}) in four dimensions and rescaling the couplings according to $g_m=\frac{1}{8\pi^2}\bar{g}_m$ we eventually arrive at
\begin{align}
 \nonumber \dot{g}_1 &= \vare g_1 -\Bigl(26+\frac{73}{3}K^2\Bigr)g_1^2-(8+8K^2)g_1g_2-\Bigl(8+\frac{14}{3}K^2\Bigr)g_2^2-\Bigl(72+\frac{226}{3}K^2\Bigr)g_1g_3-4K^2(g_2g_3+g_2g_4) \\
 \label{eps4} &-\Bigl(56+\frac{176}{3}K^2\Bigr)g_3^2-(40+42K^2)g_1g_4-(48+56K^2)g_3g_4-\Bigl(24+\frac{68}{3}K^2\Bigr)g_4^2 +\mathcal{O}(K^3),\\
 \nonumber \dot{g}_2 &= \vare g_2 -\Bigl(12+\frac{26}{3}K^2\Bigr)g_1g_2-(18+18K^2)g_2^2-\frac{1}{3}K^2g_1g_3-(4+4K^2)g_2g_3-\Bigl(2+\frac{4}{3}K^2\Bigr)g_3^2-\frac{5}{3}K^2g_1g_4\\
 \label{eps5} &-\Bigl(36+\frac{112}{3}K^2\Bigr)g_2g_4-(4+4K^2)g_3g_4-\Bigl(18+\frac{52}{3}K^2\Bigr)g_4^2+\mathcal{O}(K^3),\\
 \nonumber \dot{g}_3 &=\vare g_3 -\frac{1}{6}K^2(g_1^2+2g_2^2)-(12+9K^2)g_1g_3 +\frac{2}{3}K^2g_2g_3 -\Bigl(44+\frac{106}{3}K^2\Bigr)g_3^2+\frac{1}{3}K^2g_1g_4\\
 \label{eps6} &-(8+6K^2)g_2g_4+(8+8K^2)g_3g_4-\Bigl(8+\frac{20}{3}K^2\Bigr)g_4^2+\mathcal{O}(K^3),\\
 \nonumber \dot{g}_4 &=\vare g_4 -\frac{1}{6}K^2(g_1^2+4g_1g_2+2g_2^2)+\frac{2}{3}K^2g_1g_3-\Bigl(16+\frac{38}{3}K^2\Bigr)g_2g_3+\Bigl(12+\frac{32}{3}K^2\Bigr)g_3^2-\Bigl(12+\frac{26}{3}K^2\Bigr)g_1g_4\\
 \label{eps7} &-\Bigl(8+\frac{14}{3}K^2\Bigr)g_2g_4-(40+32K^2)g_3g_4-(8+6K^2)g_4^2+\mathcal{O}(K^3).
\end{align}
\end{widetext}

\section{Initial values for Luttinger semimetals from mean-field theory}\label{AppIni}
In this section we summarize the initial values for the bosonic RG flow that is deduced from the mean-field theory for Luttinger electrons at a three-dimensional quadratic band touching point.

\begin{figure}[t!]
\centering
\includegraphics[width=7.6cm]{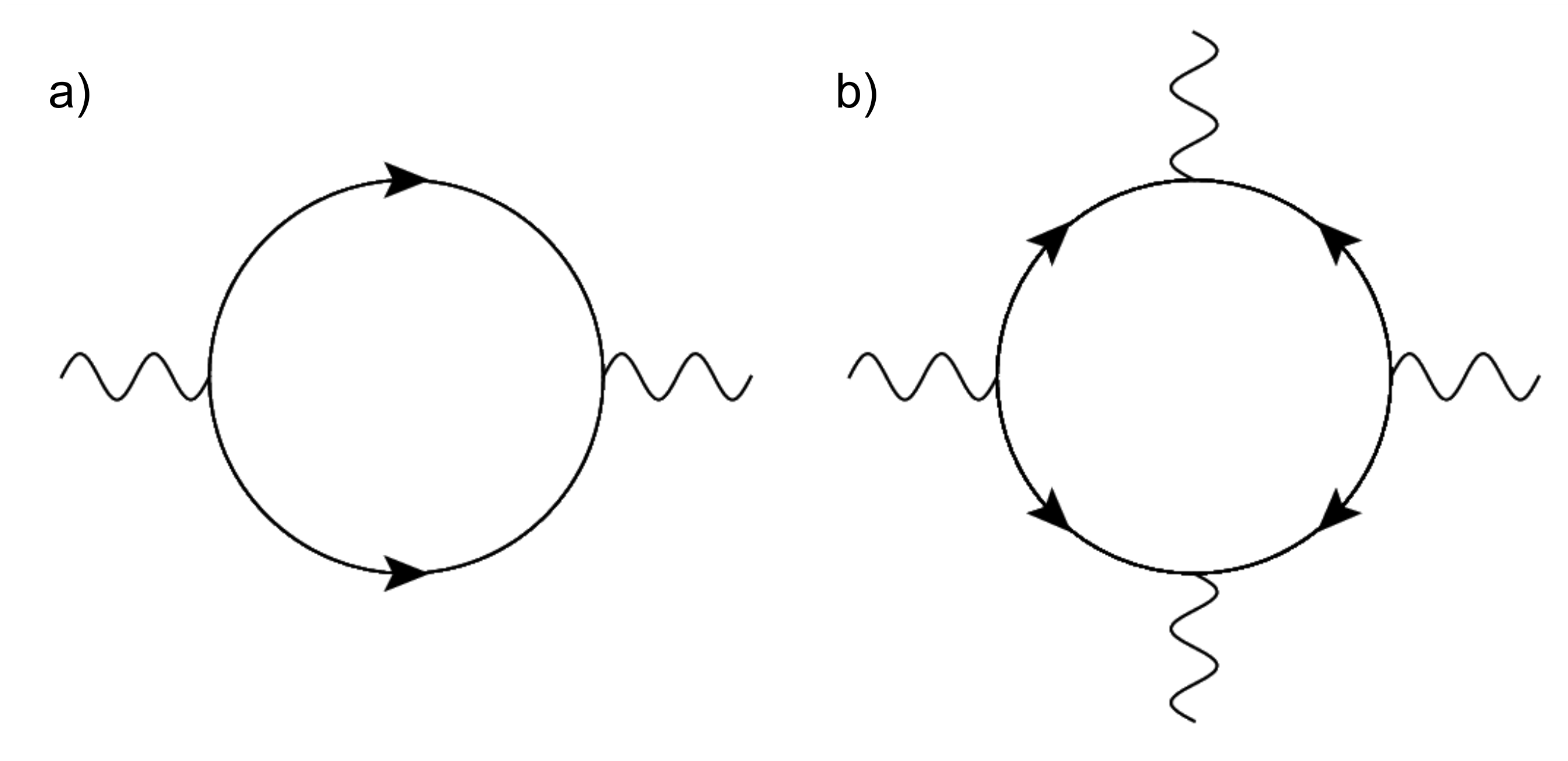}
\caption{Loop integrations that yield the mean-field expressions for the coefficients of the free energy in Eqs. (\ref{ini1})-(\ref{ini16}). Here a straight line represents a fermion propagator and a wiggly line denotes an insertion of $\Delta_a$ or $\Delta_a^*$ with vertices $\gamma_{45}\gamma_a$ \cite{BoettcherComplex}. The contribution depicted in a) results in the terms $r,a_{1,2}$ and $X,Y,Z$ that are quadratic in the boson field, whereas diagram b) generates the quartic contributions $q_{1,2}$.}
\label{FigLoops}
\end{figure}

The coefficients in Eq. (\ref{qbt3}) are determined from the fermion-fermion loop, see Fig. \ref{FigLoops} and Ref. \cite{BoettcherComplex}. After performing the angular momentum averaging we have
\begin{widetext}
\begin{align}
 \label{ini1} r={}&\frac{1}{g} -\frac{\Lambda}{10\pi^2} + \int_{\textbf{q}}^\Lambda\Bigl(T\sum_n \frac{-2q_0^2+\frac{6}{5}q^4-2\mu^2}{[q_0^2+(q^2-\mu)^2][q_0^2+(q^2+\mu)^2]}+\frac{1}{5q^2}\Bigr),\\
 \label{ini2} q_1 ={}&\int_Q^\Lambda \frac{2q_0^4+q_0^2(-\frac{36}{5}q^4+4\mu^2)+\frac{6}{7}q^8-\frac{12}{5}q^4\mu^2+2\mu^4}{[q_0^2+(q^2-\mu)^2]^2[q_0^2+(q^2+\mu)^2]^2},\\
 \label{ini3} q_2 ={}& \int_Q^\Lambda \frac{-q_0^4+q_0^2(\frac{14}{5}q^4-2\mu^2)-\frac{27}{35}q^8+2q^4\mu^2-\mu^4}{[q_0^2+(q^2-\mu)^2]^2[q_0^2+(q^2+\mu)^2]^2},\\
 \label{ini4} Z={}& -\frac{2\mu}{5} \int_Q^\Lambda \frac{5q_0^4+q_0^2(-26q^4+10\mu^2)+q^8-6q^4\mu^2+5\mu^4}{[q_0^2+(q^2-\mu)^2]^2[q_0^2+(q^2+\mu)^2]^2},\\
 \nonumber X ={}&\frac{4}{15}\int_Q^\Lambda \frac{q^2}{[q_0^2+(q^2-\mu)^2]^3[q_0^2+(q^2+\mu)^2]^3}\Bigl(25q_0^8 +q_0^6(8q^4+50\mu^2)+q_0^4(-62q^8+198q^4\mu^2+0)\\
 \label{ini5} &+q_0^2(-48q^{12}-90q^8\mu^2+252q^4\mu^4-50\mu^6)-3q^{16}+18q^{12}\mu^2-52q^8\mu^4+62q^4\mu^6-25\mu^8\Bigr),\\
 \nonumber Y ={}& \frac{16}{35}\int_Q^\Lambda \frac{q^6}{[q_0^2+(q^2-\mu)^2]^3[q_0^2+(q^2+\mu)^2]^3}\Bigl(9q_0^6+q_0^4(19q^4+9\mu^2)+q_0^2(11q^8+30q^4\mu^2-9\mu^4)\\
 \label{ini6} &+q^{12}-11q^8\mu^2+19q^4\mu^4-9\mu^6\Bigr).
\end{align}
It is a remarkable feature of the three-dimensional theory that no quartic term $\mbox{tr}(\phi^\dagger\phi\phi^\dagger\phi)\propto J_{abcd}\Delta_a^*\Delta_b\Delta_c^*\Delta_d$ is generated from the fermion loop at the mean-field level. In the presence of a small cubic anisotropy $\delta\neq 0$ the mass terms are given by $r_{\rm E}= (r+a_1)$ and $r_{\rm T}=(r+a_2)$ with 
\begin{align}
 \label{ini15} a_1 &= \delta \frac{11}{70\pi^2}\Lambda+\delta\int_{\textbf{q}}^\Lambda\Biggl[ T\sum_n \frac{\frac{4}{35}q^4\Bigl(35q_0^4+q_0^2(38q^4+42\mu^2)+3q^8-10q^4\mu^2+7\mu^4\Bigr)}{[q_0^2+(q^2-\mu)^2]^2[q_0^2+(q^2+\mu)^2]^2}-\frac{11}{35q^2}\Biggr]+\mathcal{O}(\delta^2),\\
 \label{ini16} a_2&=  -\delta\frac{1}{14\pi^2}\Lambda+\delta \int_{\textbf{q}}^\Lambda\Biggl[T\sum_n \frac{-\frac{4}{35}q^4\Bigl(-7q_0^4+q_0^2(2q^4+14\mu^2)+9q^8-30q^4\mu^2+21\mu^4\Bigr)}{[q_0^2+(q^2-\mu)^2]^2[q_0^2+(q^2+\mu)^2]^2}+\frac{1}{7q^2}\Biggr]+\mathcal{O}(\delta^2).
\end{align}
\end{widetext}
These two expressions are obtained along the lines of Ref. \cite{PhysRevB.93.205138}. The frequency and momentum integration is parametrized as
\begin{align}
 \nonumber \int_Q^\Lambda (\dots)&:= T \sum_n \int_{\textbf{q}}^\Lambda (\dots) \\
 \label{ini16b} &:= T \sum_{n=-\infty}^\infty \int_{0}^{\Lambda} \mbox{d}q q^2 \frac{1}{(2\pi)^3} \int_{\mathbb{S}^2}(\dots).
\end{align}
The frequency integration is replaced by a summation over fermionic Matsubara frequencies $q_0=\omega_n=2\pi(n+1/2)T$ due to $T>0$. This yields an infrared regularization of the expressions.

In order to make the transition from quantum to classical scaling from Eq. (\ref{qbt3}) to (\ref{qbt4}) and to read off the initial values of the boson RG, we keep a general dimension $d$, while eventually setting $d=3$. The temperature $T$ defines a reference momentum scale that we denote by $k=T^{1/2}$. The field $\Delta_a$ scales like $\Delta_a \sim T=k^2$. The remaining couplings scale like $r\sim T^{\frac{d-2}{2}}$, $q_m \sim T^{\frac{d-6}{2}}$, $X,Y,Z\sim T^{\frac{d-4}{2}}$. Consequently, we introduce dimensionless scaling functions $f$ according to
\begin{align}
 \label{ini7} r &= \frac{1}{g} -\frac{\Lambda}{10\pi^2} + T^{\frac{d-2}{2}} f_r(\mu/\Lambda^2,\mu/T),\\
 \label{ini8} q_m &= T^{\frac{d-6}{2}} f_{q_m}(\mu/\Lambda^2,\mu/T),\\
 \label{ini9} Z &=T^{\frac{d-4}{2}} f_Z(\mu/\Lambda^2,\mu/T),\\
  X & =T^{\frac{d-4}{2}} f_X(\mu/\Lambda^2,\mu/T),\\
  Y &= T^{\frac{d-4}{2}} f_Y(\mu/\Lambda^2,\mu/T).
\end{align}
For the critical theory ($r=0$), temporal fluctuations are suppressed by $T>0$ and we can neglect the $\tau$-dependence of the field $\Delta_a$. Consequently, $\int_0^{1/T}\mbox{d}\tau(\dots) = \frac{1}{T}(\dots)$ in Eq. (\ref{qbt3}), and we arrive at
\begin{align}
 \nonumber S_\Omega = \int\mbox{d}^dx \Biggl({}& T^{\frac{d-6}{2}}\Delta_a^*\Bigl[-f_X\delta_{ab}\nabla^2+\frac{f_Y}{\sqrt{3}}J_{abc}d_c(-\rmi \nabla)\Bigr]\Delta_b \\
 \label{ini10} &+T^{\frac{d-8}{2}} f_{q_1} |\vec{\Delta}|^4 +T^{\frac{d-8}{2}} f_{q_2} |\vec{\Delta}^2|^2 \Biggr).
\end{align}
For the cases considered here $f_X>0$ so that we can perform a field redefinition according to
\begin{align}
 \label{ini11} \vphi_a = T^{\frac{d-6}{4}}\sqrt{f_X} \Delta_a.
\end{align}
The action expressed in terms of $\vphi_a$ reads
\begin{align}
 \nonumber S_\Omega ={}& \int\mbox{d}^dx \Biggl( \vphi_a^*\Bigl[-\delta_{ab}\nabla^2+\frac{f_Y/f_X}{\sqrt{3}}J_{abc}d_c(-\rmi \nabla)\Bigr]\vphi_b \\
 \label{ini12} &+T^{\frac{4-d}{2}} (f_{q_1}/f_X^2) |\vec{\vphi}|^4 +T^{\frac{4-d}{2}} (f_{q_2}/f_X^2) |\vec{\vphi}^2|^2 \Biggr).
\end{align}
This action is the starting point for our classical field theory for the complex tensor boson. We identify the initial values
\begin{align}
 \label{ini13} K_\Omega &= \frac{f_Y(\mu/\Lambda^2,\mu/T)}{f_X(\mu/\Lambda^2,\mu/T)},\\
 \nonumber \lambda_{m\Omega} &= \frac{\text{S}_d\Omega^{d-4}}{(2\pi)^d} T^{\frac{4-d}{2}} \frac{f_{q_m}(\mu/\Lambda^2,\mu/T)}{f_X^2(\mu/\Lambda^2,\mu/T)}\\
 \label{ini14}&= \frac{\text{S}_d}{(2\pi)^d} \Bigl(\frac{T}{\Omega^2}\Bigr)^{\frac{4-d}{2}} \frac{f_{q_m}(\mu/\Lambda^2,\mu/T)}{f_X^2(\mu/\Lambda^2,\mu/T)}.
\end{align}
In particular, $\lambda_{3\Omega}=0$ since this coupling is not generated from the fermionic mean-field theory. We already rescaled the quartic couplings according to Eq. (\ref{field19b}). Note that the field and quartic couplings scale as $\vphi_a \sim T^{\frac{d-6}{4}}\Delta_a \sim k^{\frac{d-2}{2}}$ and $\lambda_m \sim T^{\frac{4-d}{2}}  \sim k^{4-d}$ as appropriate for a classical bosonic field theory. Note further that when identifying temperature as the ultraviolet cutoff for the bosonic theory we have $\Omega^2= T$, so that the initial values for the couplings become
\begin{align}
  \label{ini17} \lambda_{m\Omega} &= \frac{\text{S}_d}{(2\pi)^d} \frac{f_{q_m}(\mu/\Lambda^2,\mu/T)}{f_X^2(\mu/\Lambda^2,\mu/T)}.
\end{align}
We use the expressions for $K_\Omega$ and $\lambda_{m\Omega}$ in Eqs. (\ref{ini13}) and (\ref{ini17}) for the analysis in Sec. \ref{SecQBT}.

Next we discuss the influence of the anisotropy $\delta$ on the mass terms. For the strong-coupling transition we set $\mu=0$ and consider the expressions
\begin{align}
 \label{ini18} r_{\rm E} &= r + \delta \cdot T^{1/2} \bar{f}_{a_1}(T/\Lambda^2)+\mathcal{O}(\delta^2),\\
 \label{ini19} r_{\rm T} &= r +\delta \cdot T^{1/2}\bar{f}_{a_2}(T/\Lambda^2)+\mathcal{O}(\delta^2)
\end{align}
with suitably defined dimensionless functions $\bar{f}_{a_m}(T/\Lambda^2)$ extracted from Eqs. (\ref{ini15}) and (\ref{ini16}). In the region of interest $T/\Lambda^2\geq 0.29$ we find
\begin{align}
 \label{ini20} \bar{f}_{a_1}(T/\Lambda^2) > \bar{f}_{a_2}(T/\Lambda^2) >0.
\end{align}
Consequently, $r_{\rm E}<r_{\rm T}$ for small $\delta<0$, whereas $r_{\rm T}<r_{\rm E}$ for small $\delta>0$. Hence $\delta<0$ suppresses fluctuations of $\vec{\chi}$, leading to $N=2$ fluctuating complex components, whereas $\delta>0$ suppresses fluctuations of $\vec{\eta}$, leading to $N=3$ fluctuating components.

We arrive at the same conclusion for the weak-coupling transition. In this case it is sufficient to consider the logarithmic divergence of the mass terms as $y=\mu/T\to \infty$. In analogy to Eq. (\ref{ini7}) we define
\begin{align}
 \label{ini21} a_1 &= \delta \frac{11}{70\pi^2}\Lambda+ \delta \cdot T^{1/2} f_{a_1}(\mu/\Lambda^2,\mu/T),\\
 \label{ini22} a_2 &= -\delta\frac{1}{14\pi^2}\Lambda+\delta \cdot T^{1/2} f_{a_2}(\mu/\Lambda^2,\mu/T).
\end{align}
The behavior for large $y$ is then given by
\begin{align}
 \label{ini23} f_r(\mu/\Lambda^2,y) & \to  \frac{1}{10\pi^2}(-0.30-0.5\log y) y^{1/2},\\
  \label{ini24} f_{a_1}(\mu/\Lambda^2,y) & \to  \frac{1}{10\pi^2}(-1.54+0.75\log y) y^{1/2},\\
 \label{ini25} f_{a_2}(\mu/\Lambda^2,y) & \to  \frac{1}{10\pi^2}(1.18-0.25\log y) y^{1/2}
\end{align}
as $y\to \infty$. The divergence of $\log y$ dominates all other contributions and the corrections to the mass terms can be deduced from the sign of $\log(y)y^{1/2}$ on the right-hand sides of Eqs. (\ref{ini23})-(\ref{ini25}). We observe that again $r_{\rm E}$ is reduced for small $\delta<0$, whereas $r_{\rm T}$ is reduced for $\delta>0$.

\end{appendix}

\bibliography{refs_fluct}

\end{document}